\documentclass[trackchanges]{aastex701}

\usepackage{graphicx}       
\usepackage[percent]{overpic} 
\usepackage{subcaption}     
\usepackage{adjustbox}      
\usepackage{nameref}        

\begin{document}

\title{Machine Learning-based Separation of the $\text{He~\sc{i}}~10830~\text{\AA}$ \ Chromospheric Signal: Quantitative Analysis of Chromosphere-Corona Intensity in the Quiet Sun}

\author[orcid=0009-0003-8982-6027,sname='Li']{Huaiming Li}
\affiliation{Yunnan Observatories, Chinese Academy of Sciences, Kunming 650216, People’s Republic of China}
\affiliation{University of Chinese Academy of Sciences, Beijing 101408, People’s Republic of China}
\email{lihuaiming23@mails.ucas.ac.cn}  

\author{Fangyu Xu} 
\correspondingauthor{Fangyu Xu} 
\email{xu\_fangyu@ynao.ac.cn}
\affiliation{Yunnan Observatories, Chinese Academy of Sciences, Kunming 650216, People’s Republic of China}
\email{}

\author[0000-0002-5302-3404]{Yi Bi} 
\affiliation{Yunnan Observatories, Chinese Academy of Sciences, Kunming 650216, People’s Republic of China}
\email{}

\author[0000-0001-7575-5449]{Zhenyu Jin}
\affiliation{Yunnan Observatories, Chinese Academy of Sciences, Kunming 650216, People’s Republic of China}
\affiliation{Yunnan Key Laboratory of Solar Physics and Space Science, 650216, People’s Republic of China}
\email{}

\begin{abstract}
The $\text{He~\textsc{i}}~10830~\text{\AA}$ line, a crucial optically thin chromospheric line, is frequently used to study coronal heating and vertical coupling across the chromosphere-corona interface. However, its images are severely contaminated by the strong photospheric background signal, hindering the analysis of fine chromospheric structures.
Given the morphological differences between the Active Region ($\text{AR}$) and the Quiet Sun ($\text{QS}$), we proposed separating the $\text{He~\textsc{i}}~10830~\text{\AA}$ chromospheric signal using two deep learning $\text{CNN}$ models.
Our model utilizes $\text{TiO}$ images and cross-band learning to infer the $\text{He~\textsc{i}}~10830~\text{\AA}$ photospheric background. The output is combined with an exponential absorption model to achieve quantitative analysis of the pure chromospheric component.
Joint analysis of Solar Dynamics Observatory ($\text{SDO}$) data and the separated $\text{QS}$ structures reveals a strong spatial negative correlation between chromospheric $\text{He~\textsc{i}}~10830~\text{\AA}$ intensities($\text{R} \approx -0.84$ in $304\,\text{\AA}$ ), and significant layered coupling with $\text{EUV}$ (171, 193, and 304 \AA) radiation.
Furthermore, strong $\text{He~\textsc{i}}~10830~\text{\AA}$ absorption areas are highly correlated with regions of strong magnetic fields, while $\text{171\,\text{\AA}}$ radiative enhancement areas extend to the strong magnetic field edges and the mixed-polarity regions.
These findings quantify the radiation intensity relationship between $\text{He~\textsc{i}}~10830~\text{\AA}$ and $\text{EUV}$ bands in the Quiet Sun.
It also demonstrates the differences in heating characteristics between unipolar and mixed-polarity magnetic fields.

\end{abstract}
\keywords{\uat{Solar chromosphere}{1479} --- \uat{Quiet sun}{1322} --- \uat{Active solar regions}{1974} --- \uat{Solar coronal heating}{1989} }

\section{Introduction}\label{sec:Introduction}
The vertical coupling and associated observations across the entire solar atmosphere, from the photosphere through the chromosphere to the corona, have long been an important unsolved mystery. Understanding this coupling necessitates fine observations of the interface, which is characterized by extremely steep temperature and density gradients.
The $\text{He~\textsc{i}}~10830~\text{\AA}$ triplet spectral line, an optically thin band formed in the chromosphere–corona interface, carries multi-layered atmospheric information and is crucial for studying energy and mass coupling between different solar atmospheric layers. 
Its absorption primarily arises from the $2\text{s}\,^3\text{S} \to 2\text{p}\,^3\text{P}$ transition in Orthohelium, forming a triplet at 10829.09, 10830.25, and 10830.34\,\AA\ with a theoretical intensity ratio of 1:3:5 \citep{Cite0_3}, while the $2\text{s}\,^3\text{S}$ metastable state requires high energy for excitation \citep{Cite0_1}.
One of the primary excitation mechanisms for $\text{He~\textsc{i}}~10830~\text{\AA}$ is the photoionization-recombination (PR) process \citep{Cite0_4}.
The line is highly sensitive to local heating, making it a valuable diagnostic of energy release and vertical coupling across the chromosphere-to-corona layers \citep{Cite0_2}.
Due to its relatively high formation temperature, the $\text{He~\textsc{i}}~10830~\text{\AA}$ line is representative of chromospheric heating. Specifically, the PR mechanism remains the dominant formation process below 20,000 K, with a characteristic temperature typically modeled at 10,000 K \citep{Cite0_5, Cite0_6}.
The line is crucial for studying the structure and evolution of the transition region, tracing the photospheric origin of upper atmospheric activity, and the energy and mass coupling between different solar layers.

Earlier studies using $\text{He~\textsc{i}}~10830~\text{\AA}$ data focused on coronal heating observation. These studies utilized multi-layer $\text{He~\textsc{i}}~10830~\text{\AA}$ combined with $\text{SDO}$ satellite data (including the Atmospheric Imaging Assembly ($\text{AIA}$) and the Helioseismic and Magnetic Imager ($\text{HMI}$)) \citep{Cite15, Cite16, Cite17}, alongside fine-channel coronal heating observations from the Goode Solar Telescope ($\text{GST}$) \citep{Cite18, Cite1}. 
Moreover, related research also focuses on the photospheric origins of coronal heating, which is investigated through high-resolution observations of small-scale dynamics such as: granule-driven magnetic reconnection \citep{Cite2}, granule-driven quasi-periodic microjets \citep{Cite6}, and the structure of various chromospheric jets \citep{Cite8, Cite3, Cite7, Cite9}.  
Furthermore, Swedish Solar Telescope ($\text{SST}$) raster scan data has been specifically used for observations of Ellerman bombs \citep{Cite27, Cite13}. Advanced techniques, including polarization observations for flow velocities and $\text{Fabry-Pérot}$ multi-band scanning \citep{Cite14, Cite12}, underscore the line's unique value in tracing magnetically controlled energy dissipation.

Previous studies of the $\text{He~\textsc{i}}~10830~\text{\AA}$ line, particularly observations of the Moss regions \citep{Cite19, Cite20}, were primarily conducted based on data mixing photospheric and chromospheric signals and analyzed using the empirical threshold method (e.g., setting $90\%$ of the mean $\text{He~\textsc{i}}~10830~\text{\AA}$ intensity as a threshold) to track the photospheric origin of coronal heating and its relation to the magnetic field\citep{Cite4, Cite5, Cite11}. 
However, the strong photospheric signal in $\text{He~\textsc{i}}~10830~\text{\AA}$ imaging severely limits the precise separation and quantitative analysis of chromospheric and photospheric contributions, hindering the study of fine chromospheric structures.

To address the strong photospheric interference, we developed a novel deep learning framework to separate the $\text{He~\textsc{i}}~10830~\text{\AA}$ chromospheric signal. 
Given the fundamentally different structural characteristics between $\text{AR}$ and the$\text{QS}$ in the $\text{He~\textsc{i}}~10830~\text{\AA}$ band, we propose two independent deep Convolutional Neural Network (CNN) models based on the $\text{ResNet}$ architecture \citep{Cite22}. 
These models utilize temporally and spatially matched $\text{TiO}$ images as input. The spatial matching is achieved using algorithms such as $\text{Scale-Invariant Feature Transform}$ ($\text{SIFT}$) and $\text{Optical Flow}$\citep{Cite24, Cite25, Cite26}, employing a cross-band learning paradigm to infer the photospheric signal in the $\text{He~\textsc{i}}~10830~\text{\AA}$ imaging.
Notably, in Model B (developed for the Quiet Sun), we incorporated Fourier Features Let Networks ($\text{FFLN}$) \citep{Cite23} to enhance the fitting capability for complex fine structures. 
Ultimately, we successfully separated the true {$\text{He~\textsc{i}}~10830~\text{\AA}$} chromospheric information for the first time using the method described in Section~\ref{subsec:Principle}.
The completeness and continuity of the separated chromospheric structures validate the effectiveness of our model and the accuracy of the information separation. This method breaks the limitations of traditional thresholding, allowing us for the first time to conduct accurate pixel-level spatial and quantitative studies on the $\text{He~\textsc{i}}~10830~\text{\AA}$ chromospheric line and the $\text{AIA}$ $\text{EUV}$ lines.


Unlike previous studies, which mostly focused on Moss regions or areas of intense magnetic fields, this work utilizes more generally representative $\text{QS}$ data. We conducted the first pixel-level joint analysis of the separated $\text{He~\textsc{i}}~10830~\text{\AA}$ fine chromospheric structures with AIA's 171\,\AA, 193\,\AA, 304\,\AA\ channels, and HMI's $I_{\text{c}}$ and $I_{\text{m}}$ data. Our analysis reveals a significant negative correlation between the $\text{He~\textsc{i}}~10830~\text{\AA}$ chromospheric intensity and the corresponding coronal intensity in the $\text{QS}$.
Furthermore, the observed spatial and layered coupling provides strong evidence for a low-atmosphere-driven mechanism for coronal heating in $\text{QS}$ fine structures. We also find that while strong $\text{He~\textsc{i}}~10830~\text{\AA}$ absorption generally correspond to intense magnetic fields, the strong $\text{AIA}$ radiative enhancement extends to the edges of the intense magnetic fields and the magnetic neutral line regions. This observation confirms that there is a spatial offset in the magnetic environment between the heating footpoints of the chromosphere and corona, and that the coronal heating mechanisms differ between mixed-polarity regions and unipolar regions in the quiet Sun.
The remainder of this paper is organized as follows: Section~\ref{sec:Data} introduces the observational data used. Section~\ref{sec:Methodology} details the data pre-processing steps, the physical principle of chromospheric signal separation, the deep learning approach, and the results. Section~\ref{sec:Results} presents the main findings from the pixel-level joint analysis. Finally, Section~\ref{sec:Discussion} provides an in-depth discussion of these results, and Section~\ref{sec:Conclusion} summarizes our primary discoveries.





\section{Instrument and Data}\label{sec:Data}

The data for this study primarily originated from two observation facilities: $\text{NVST}$ and $\text{SDO}$. $\text{NVST}$ provided high-resolution imaging data in the $\text{TiO}$ band and the $\text{He~\textsc{i}}~10830~\text{\AA}$ band, which were used for model training and subsequent chromospheric information separation.
Although not strictly hardware-synchronized, the observations are effectively simultaneous, as the 30~s interval is much shorter than the typical 5--10~min granulation timescale.
$\text{SDO}$ data, including $\text{AIA}$ multi-band imaging at 171\,\AA\ (Fe~\textsc{ix}, $\log T \approx 5.8$), 193\,\AA\ (Fe~\textsc{xii}, $\log T \approx 6.1$; Fe~\textsc{xxiv}, $\log T \approx 7.3$), and 304\,\AA\ (He~\textsc{ii}, $\log T \approx 4.7$) channels \citep{Cite16} , and HMI's $I_{\text{c}}$ and $I_{\text{m}}$ data,referring to continuum intensity and line-of-sight magnetograms respectively \citep{Cite17}, were used for physical interpretation and analysis.
The NVST is located at Fuxian Lake, Yunnan \citep{Cite21}. Its \text{He~\textsc{i}}~10830~\text{\AA} channel achieved first light in 2022 and was upgraded in 2023 to a $1024 \times 1280$ pixel scientific-grade detector for routine scientific observations \citep{Cite10}.
This high-performance near-infrared camera was developed through a collaboration between the astronomical technology laboratory at Yunnan Observatories and the Kunming Institute of Physics.
The observation line center of this channel is located at 10830.35\,\AA, with a FWHM of 0.5\,\AA. It is equipped with a bandpass filter featuring a tunable range of $\pm$5\,\AA.
High-resolution image reconstruction was performed using the Non-rigid Alignment (NASIR) method \citep{NASIR} at Yunnan Observatories, Chinese Academy of Sciences.

For model training, two independent datasets were utilized. The AR model employed TiO images as input and the \text{He~\textsc{i}}~10830.35~\text{\AA} blue-shifted 1.00\,\AA\ off-band images as the photospheric label, collected over multiple
days. 
For brevity, \text{He~\textsc{i}}~10830.35~\text{\AA}  and its blue-shifted counterpart at 10829.35\text{\AA} are hereafter referred to as 10830 \text{\AA} and 10829 \text{\AA}, respectively.
The $10829.35\,\mathrm{\AA}$ off-band signal is primarily dominated by the photospheric continuum, with chromospheric contributions effectively suppressed by four factors. 
First, the He~\textsc{i} $10829.09\,\mathrm{\AA}$ line intensity is only $1/8$ of the $10830.34\,\mathrm{\AA}$ main peak \citep{Cite0_3}. Second, our filter is centered $0.26\,\mathrm{\AA}$ away from this weak line core, placing it outside the FWHM range. 
Additionally, observed morphological features (e.g., granulation and sunspots) are identical to pure photospheric structures, suggesting any residual chromospheric signal is reduced to noise levels. 
Finally, the deep learning model naturally treats spatially uncorrelated chromospheric components as random residuals. 
Collectively, these mechanisms ensure this band provides a stable and reliable photospheric background.
Chromospheric signals were extracted from observations on July 16, 2024. For the QS model, a single-day dataset (November 19, 2024, 05:20–06:10 UT) was used for training. Because the chromospheric signal in the QS is weak, this approach is well-suited for residual-optimization-like algorithms, enabling more accurate extraction of fine-scale structures.
To ensure the quality of model training and the subsequent chromospheric structure separation, we selected datasets with favorable seeing conditions. The data were split into training and test sets at an $85\%$-$15\%$ ratio, using a random seed of $42$ for reproducibility. Key observational parameters regarding spatial and temporal resolution are summarized in Table~\ref{tab:1}.
\begin{deluxetable}{lccccc}
\tablewidth{0pt} 
\tablecaption{Summary of Key Observational Data Properties \label{tab:1}}
\tablehead{
\colhead{Channel} & \colhead{TiO} & \colhead{He I} & \colhead{HMI} & \colhead{AIA}
}
\startdata
Data & 7058 \text{\AA} & 10830 \text{\AA} & I\textsubscript{c} , I\textsubscript{m} & 171, 193, 304 \text{\AA} \\
Temporal Cadence & $30\,\text{s}$ & $8\,\text{s}$ & $45\,\text{s}$ & $12\,\text{s}$ \\
Spatial Resolution & $0.052\arcsec$ & $0.093\arcsec$ & $0.6\arcsec$ & 0.6\arcsec \\
\enddata

\end{deluxetable}


\section{Methodology}\label{sec:Methodology}

\subsection{Pre-processing}  
The entire research process, from data acquisition to the final separation of chromospheric information, is divided into five sequential stages: 
Firstly, high-resolution observational data were acquired from $\text{NVST}$ and reconstructed.
Secondly, for the time-matched images, precise spatial registration was achieved by employing SIFT \citep{Cite24, Cite25} and the Optical Flow method \citep{Cite26}. The \text{He~\textsc{i}}~10830~\text{\AA} data was utilized as the reference frame for this process.
Thirdly, two ResNet-based neural network models were trained to learn the mapping relationship between the TiO images and the corresponding \text{He~\textsc{i}}~10830~\text{\AA} band photospheric signal.
Fourthly, the trained models were used to predict the photospheric equivalent signal in the \text{He~\textsc{i}}~10830~\text{\AA} band, providing a reference for distinguishing the photospheric and chromospheric components. 
Finally, we successfully extracted the \text{He~\textsc{i}}~10830~\text{\AA} chromospheric signal from the \text{He~\textsc{i}}~10830~\text{\AA} reconstructed data.


\subsection{Physical Principle and Motivation}  \label{subsec:Principle}
We propose a dual-model $\text{ResNet}$-based machine learning approach aimed at processing the complex and highly variable chromospheric signal in the \text{He~\textsc{i}}~10830~\text{\AA} band data under a strong photospheric background.
This complexity fundamentally arises from the superposition of intense photospheric signals with weak chromospheric signals. The physical origins of this phenomenon are multifaceted: 
firstly, the radiative transfer process of the \text{He~\textsc{i}}~10830~\text{\AA} line is optically thin, enabling the observed data to contain multi-layered information from the photosphere up to the upper atmosphere; secondly, the \text{He~\textsc{i}}~10830~\text{\AA} absorption feature originates from a difficult-to-excite, high-lying metastable state transition, resulting in a weak signal; finally, the filter with a $\text{FWHM} = 0.5\,\text{\AA}$ introduces significant photospheric continuum contamination; a narrower bandwidth would reduce the contribution of this photospheric signal.
Collectively, these three factors give rise to a low signal-to-noise ratio (SNR) of the observed data.
The main objective of this study is precisely to overcome these challenges by accurately separating and extracting the chromospheric signal in the presence of a strong photospheric background. 
Specifically, the machine learning model uses the \text{TiO} image as input and, through cross-band learning, infers the corresponding photospheric information within the \text{He~\textsc{i}}~10830~\text{\AA} band. 
The final chromospheric component ($\text{C}$) is obtained by dividing the observed signal ($\text{I}_{\text{obs}}$) by the predicted photospheric background ($\text{I}_{\text{cont}}$), a concept derived from the most fundamental exponential absorption model:
$$C = I_{\text{obs}} / I_{\text{cont}}$$
where $I_{\text{obs}}$ represents the observed \text{He~\textsc{i}}~10830~\text{\AA} signal, and $I_{\text{cont}}$ denotes the predicted photospheric signal.
Physically, this approach is a simplification of the radiative transfer model under non-LTE conditions. Since the observed signal ($I_{\text{obs}}$) is inherently a combined effect of all physical processes, we do not explicitly calculate complex processes such as photon absorption, photoionization-recombination (PR) mechanisms, scattering, or the population of He~\textsc{i} triplet states. Instead, our model focuses on reconstructing the pure photospheric background ($I_{\text{cont}}$) as a reference. Consequently, the ratio $C = I_{\text{obs}} / I_{\text{cont}}$ naturally captures and integrates all modulation effects from the chromosphere.


\begin{figure}[!htbp] 
    \centering

    \begin{minipage}{0.22\textwidth}
        \centering
        \begin{overpic}[width=\textwidth]{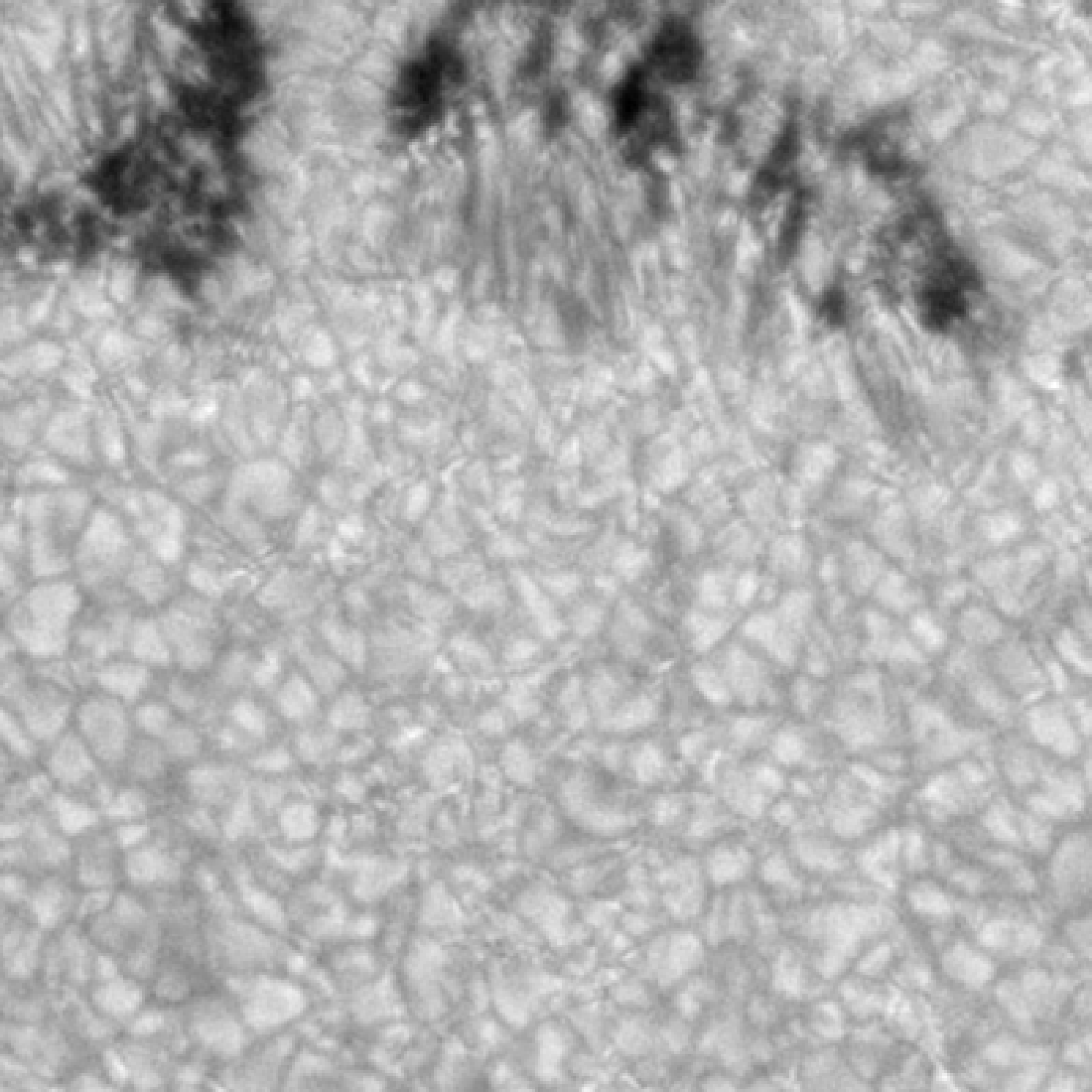}
            \put(45,5){\color{red}\textbf{(A1)}} 
        \end{overpic}
    \end{minipage}\hspace{0.3em}%
    \begin{minipage}{0.22\textwidth}
        \centering
        \begin{overpic}[width=\textwidth]{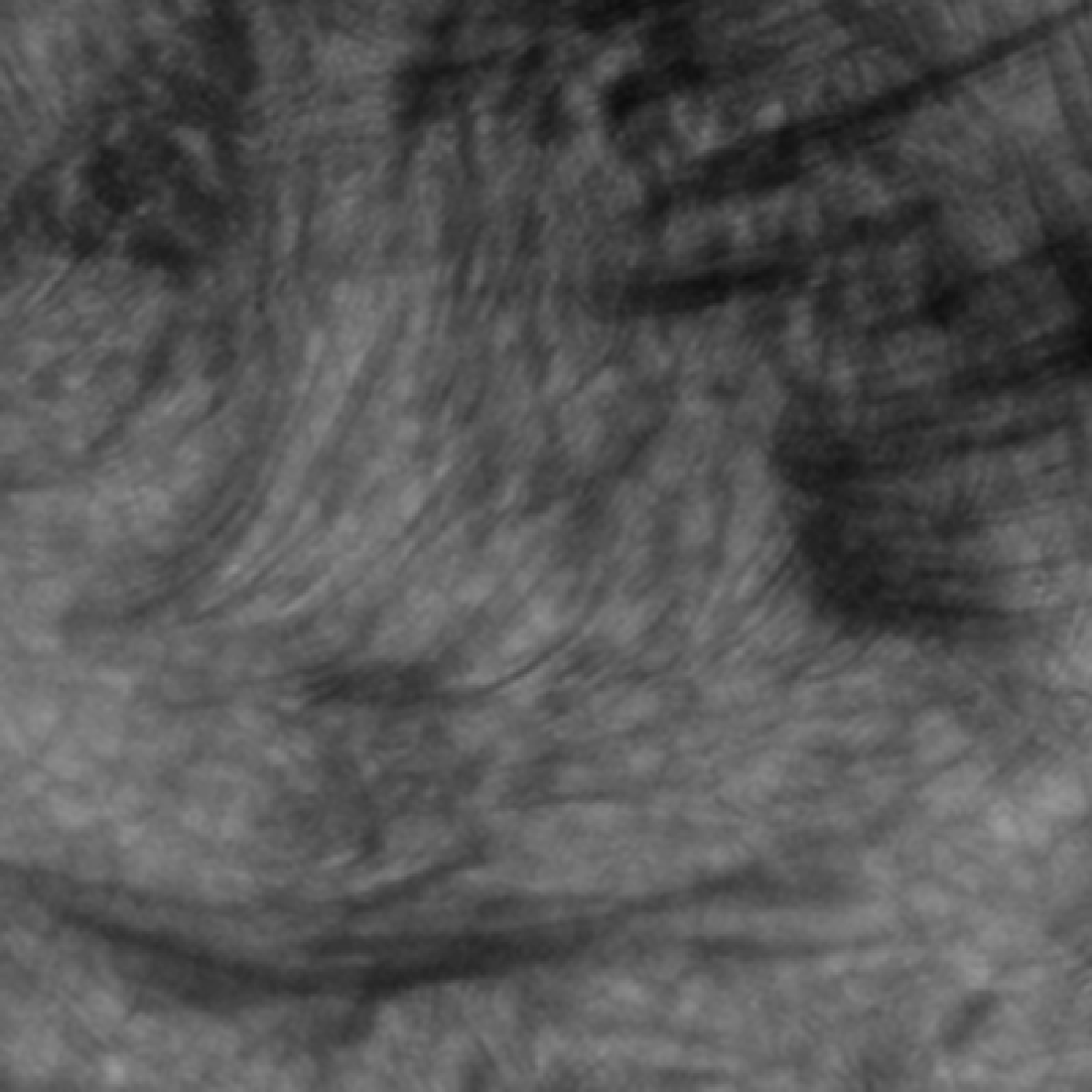}
            \put(45,5){\color{red}\textbf{(A2)}} 
        \end{overpic}
    \end{minipage}

    \vspace{3pt} 

    \begin{minipage}{0.22\textwidth}
        \centering
        \begin{overpic}[width=\textwidth]{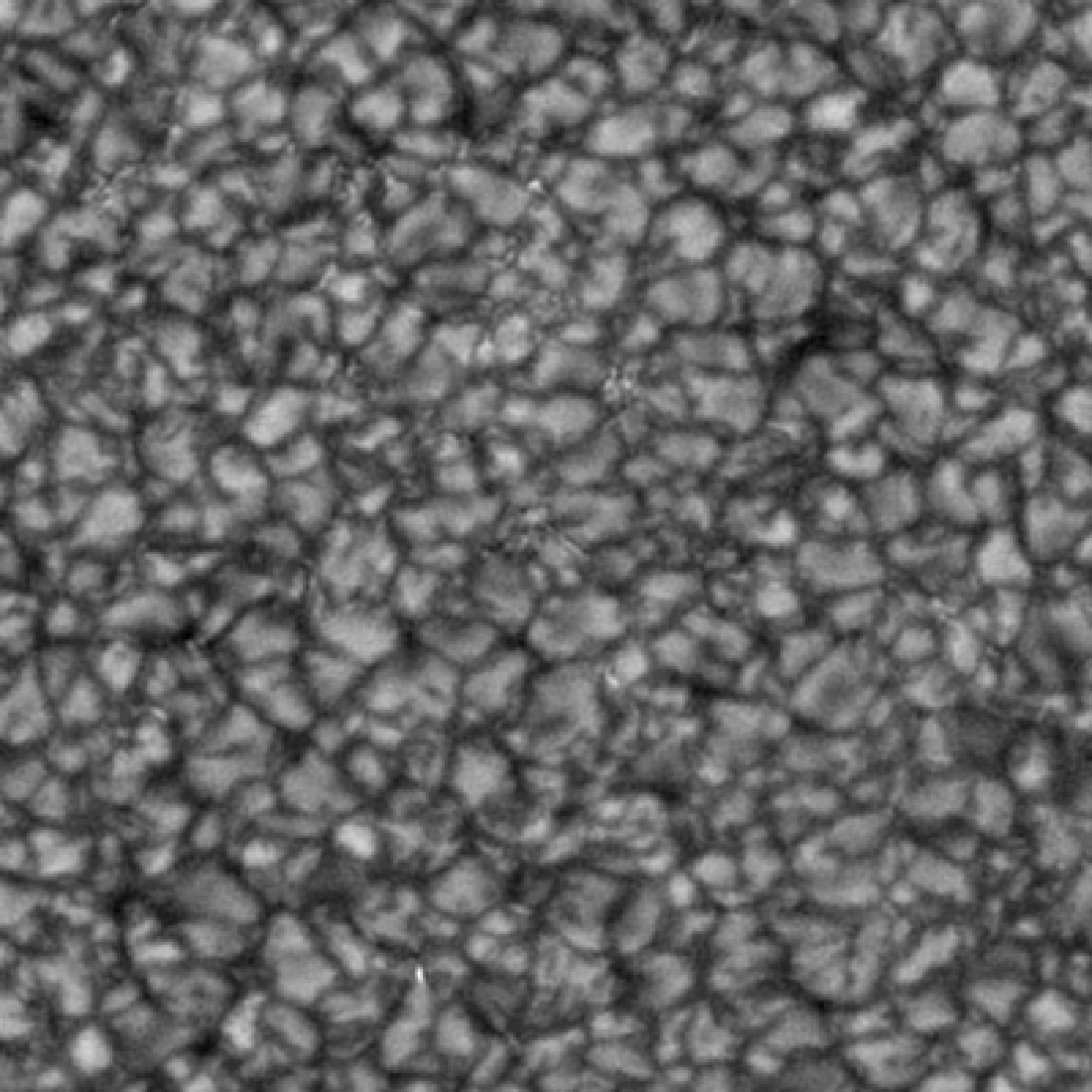}
            \put(45,5){\color{red}\textbf{(B1)}} 
        \end{overpic}
    \end{minipage}\hspace{0.3em}%
    \begin{minipage}{0.22\textwidth}
        \centering
        \begin{overpic}[width=\textwidth]{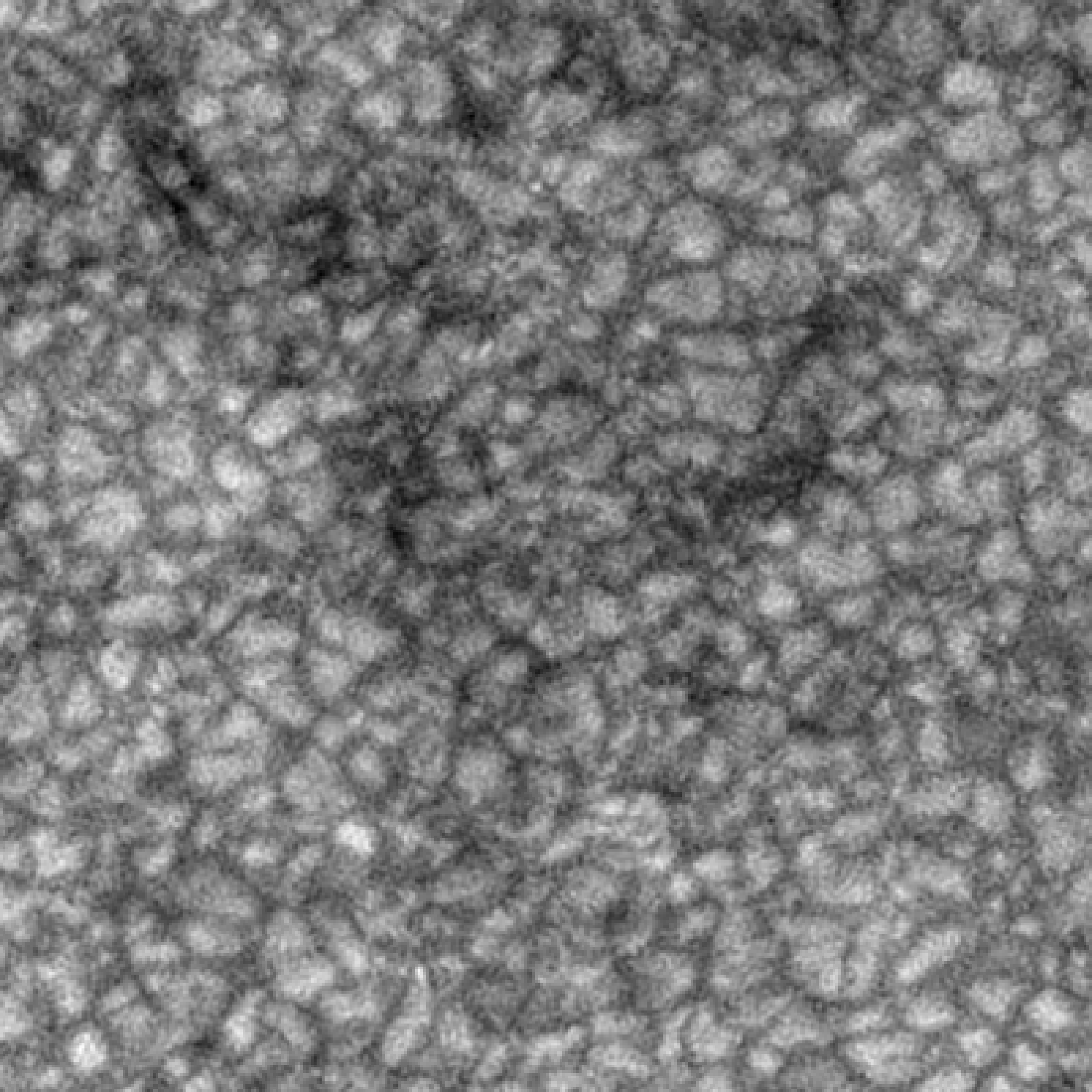}
            \put(45,5){\color{red}\textbf{(B2)}} 
        \end{overpic}
    \end{minipage}

    \caption{The first row (A1, A2) displays the high-contrast features of the Active Region dataset. 
    The second row (B1, B2) shows the low-contrast, fine-scale structures of the Quiet Sun dataset. 
    Panels (A1, B1) are the $\text{TiO}$ photospheric signals, and Panels (A2, B2) are the observed \text{He~\textsc{i}}~10830~\text{\AA} signals. All images have a spatial size of $400\times400$ pixels.}
    \label{fig:input_output}
\end{figure}

As shown in Figure~\ref{fig:input_output}, there exist substantial morphological and radiative differences between the AR and the QS, where the chromospheric signal in the AR (Panels A1-A2) is significantly stronger than the faint features in the QS (Panels B1-B2). 
To address these inherent disparities, a dual-model approach is thus adopted in this study.
The physical foundation of our method is based on the different characteristics of solar regions. Ideally, the training labels should show the 10830 \text{\AA} photosphere with as little chromospheric signal as possible. 
For Model A (Active Regions), the strong chromospheric absorption at \text{He~\textsc{i}}~10830~\text{\AA} makes it necessary to use a nearby off-band (such as 10829 $\text{\AA}$) as a reference.
This is justified because the chromospheric signal at 10829 \text{\AA} is very weak, while its photospheric intensity effectively represents the 10830 \text{\AA} background. 
For Model B (Quiet Sun), we establish a direct mapping from TiO to 10830 \text{\AA}. This is based on the assumption that chromospheric signals in quiet regions are not only weak but also have different structures compared to the photosphere. Therefore, these chromospheric features do not interfere with the model’s ability to learn the stable mapping from TiO structures to the 10830 \text{\AA} photospheric background. Table~\ref{tab:physics} summarizes the physical principles of these two models.

\begin{deluxetable}{lccccL} 
\tablewidth{0pt}
\tablecaption{Physical Principles of Model A and Model B\label{tab:physics}}
\tablehead{
\colhead{Model} & \colhead{Mapping Relation} & \colhead{Region} & \colhead{Chromospheric Signal} & \colhead{Photospheric Label}
}
\startdata
Model A & TiO $\rightarrow$ 10830 \text{\AA} Photosphere  & AR &  Strong  & 10829 \text{\AA}   \\
Model B & TiO $\rightarrow$ 10830 \text{\AA} Photosphere  & QS &  Weak & 10830 \text{\AA}  \\
\enddata

\end{deluxetable}



\begin{figure*}[!htbp]
    \centering

    \begin{minipage}{0.195\textwidth}
        \centering
        \begin{overpic}[width=\textwidth]{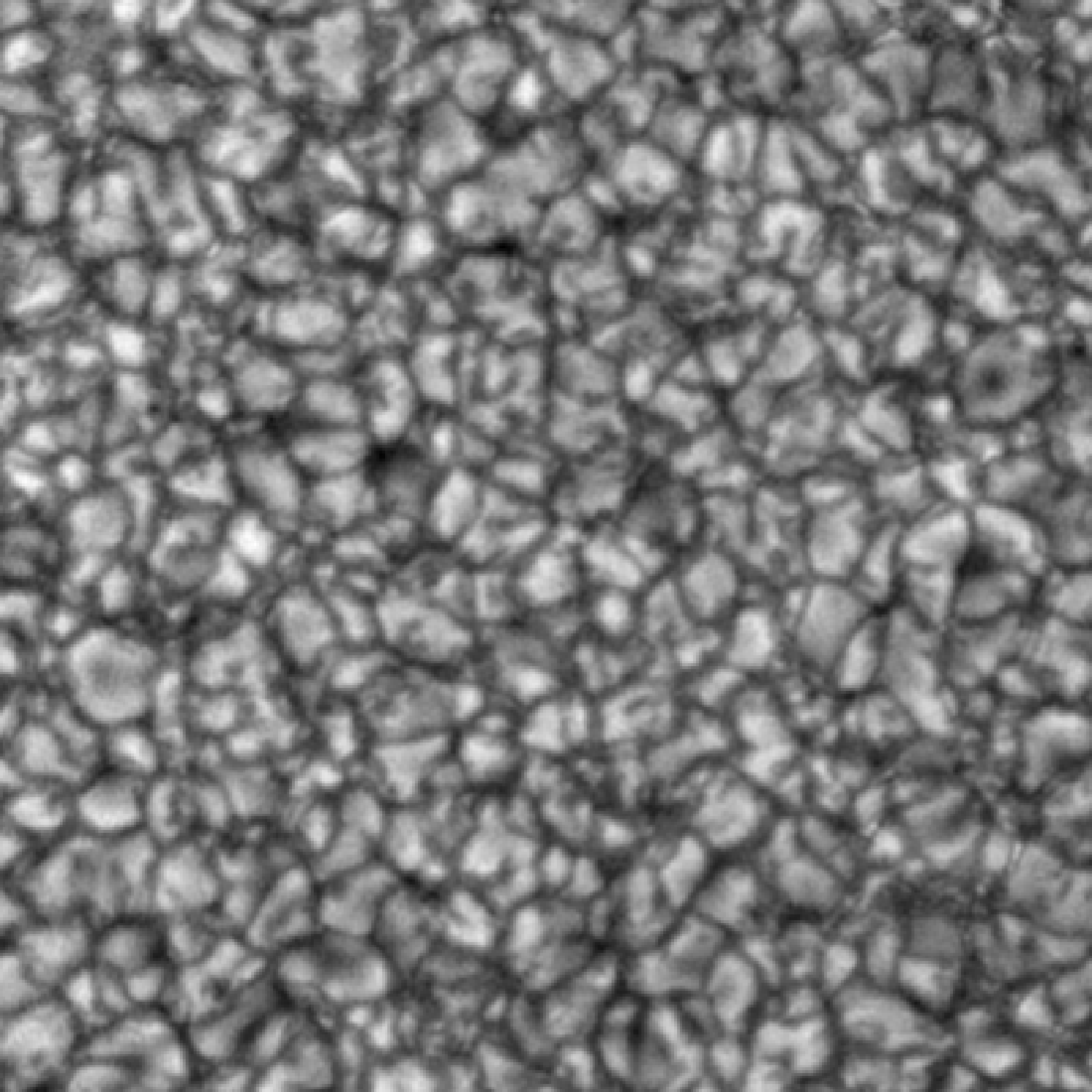}
            \put(40,5){\color{red}\large\textbf{TiO}}
        \end{overpic}
    \end{minipage}\hspace{0.2em}%
    \begin{minipage}{0.195\textwidth}\centering\includegraphics[width=\textwidth]{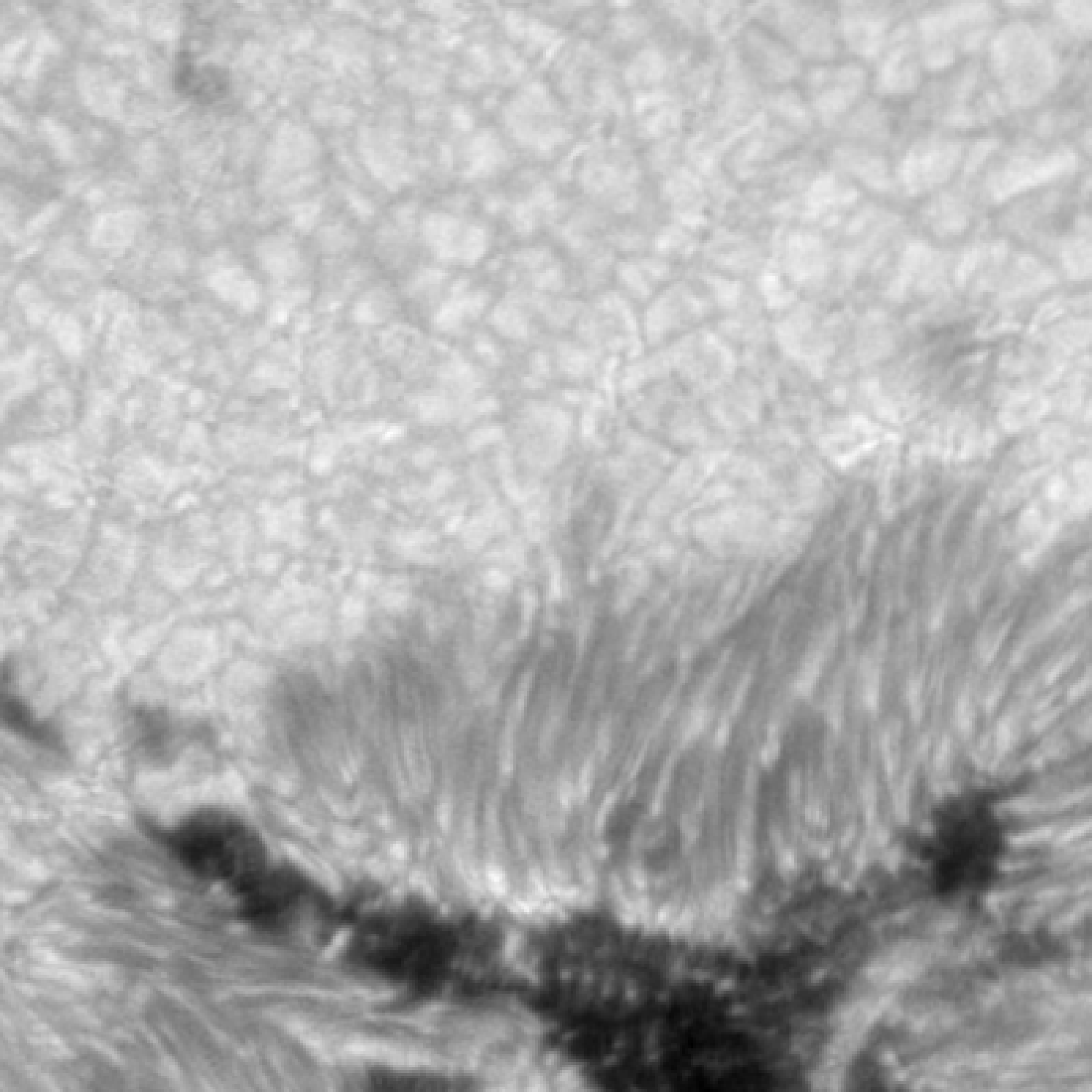}\end{minipage}\hspace{0.2em}%
    \begin{minipage}{0.195\textwidth}\centering\includegraphics[width=\textwidth]{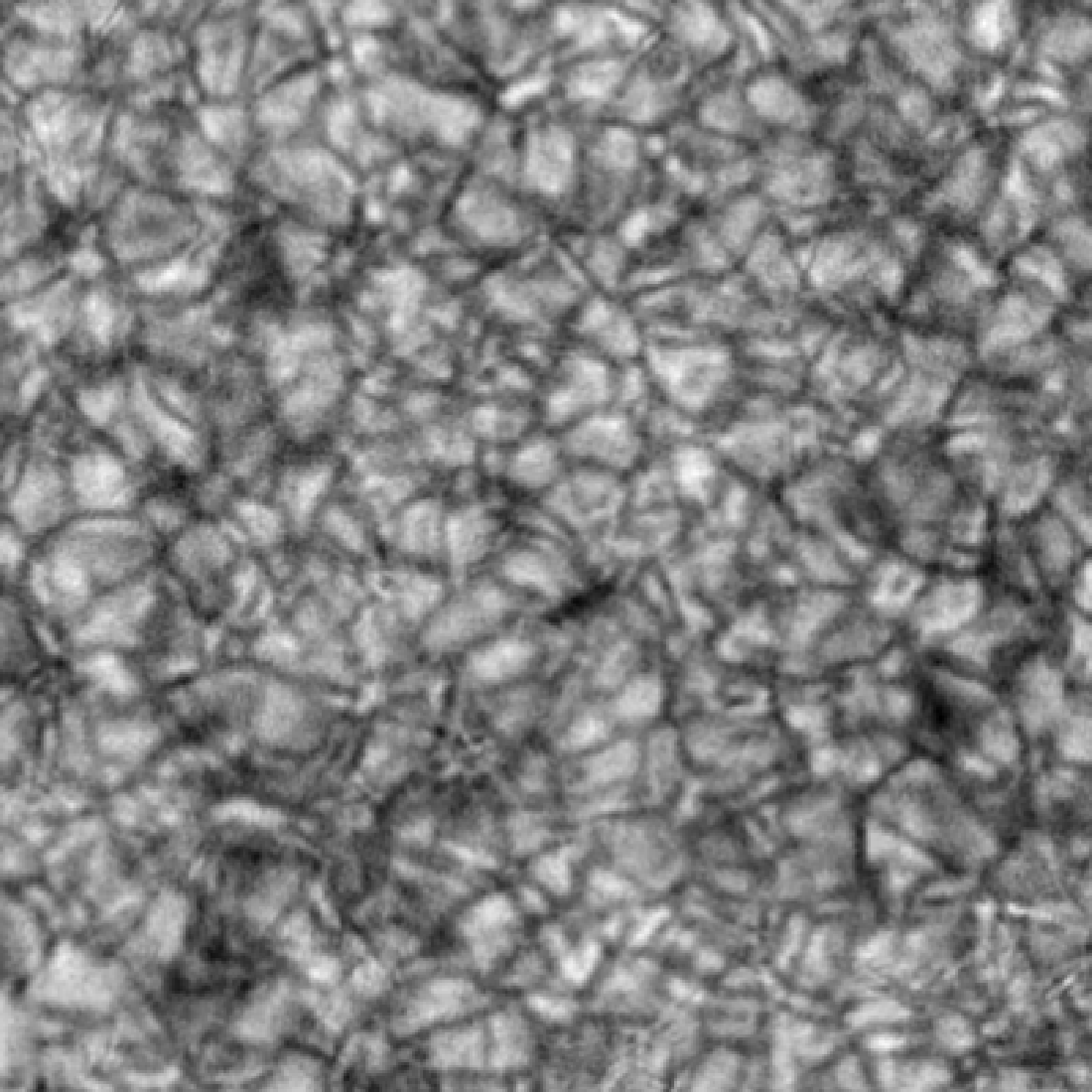}\end{minipage}\hspace{0.2em}%
    \begin{minipage}{0.195\textwidth}\centering\includegraphics[width=\textwidth]{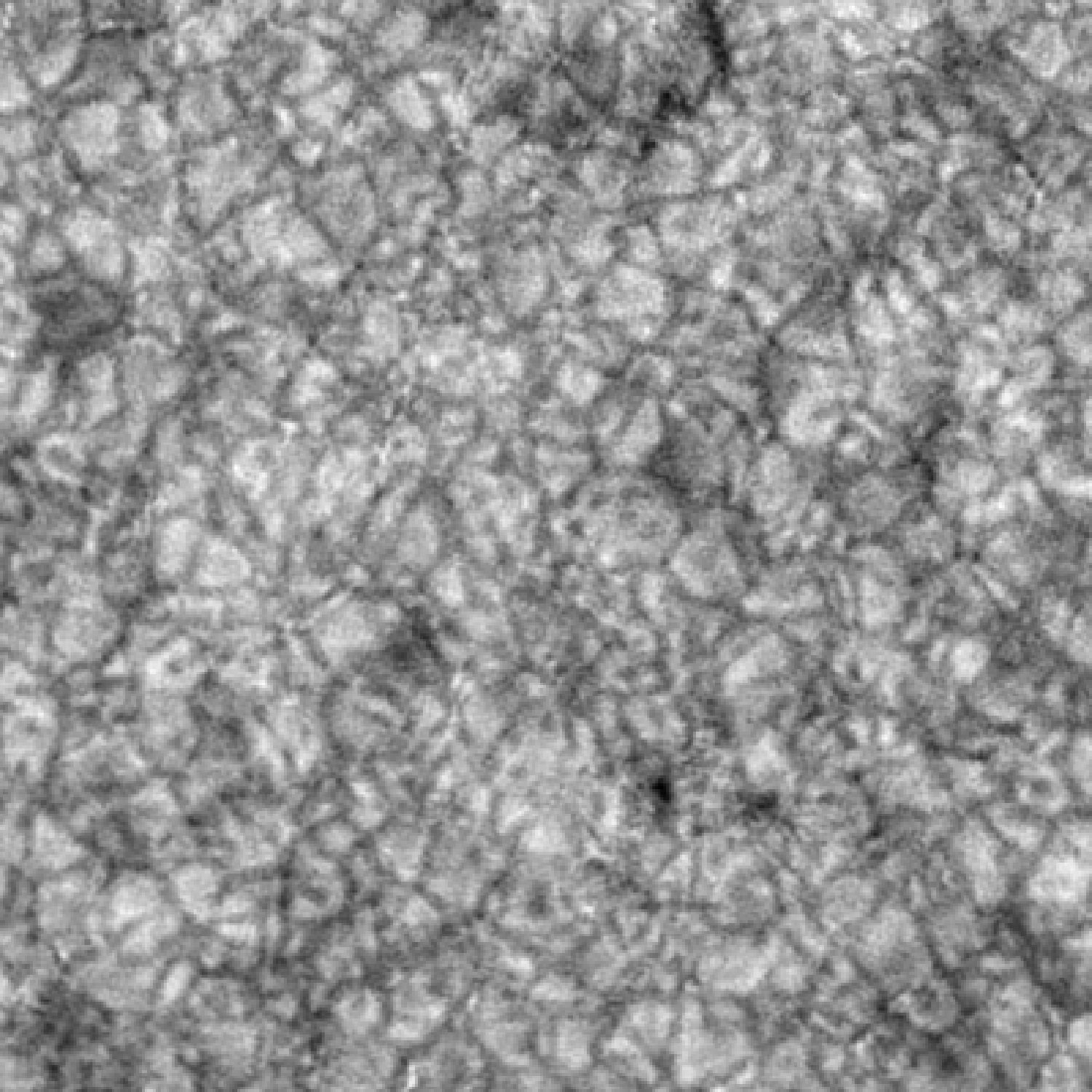}\end{minipage}\hspace{0.2em}%
    \begin{minipage}{0.195\textwidth}\centering\includegraphics[width=\textwidth]{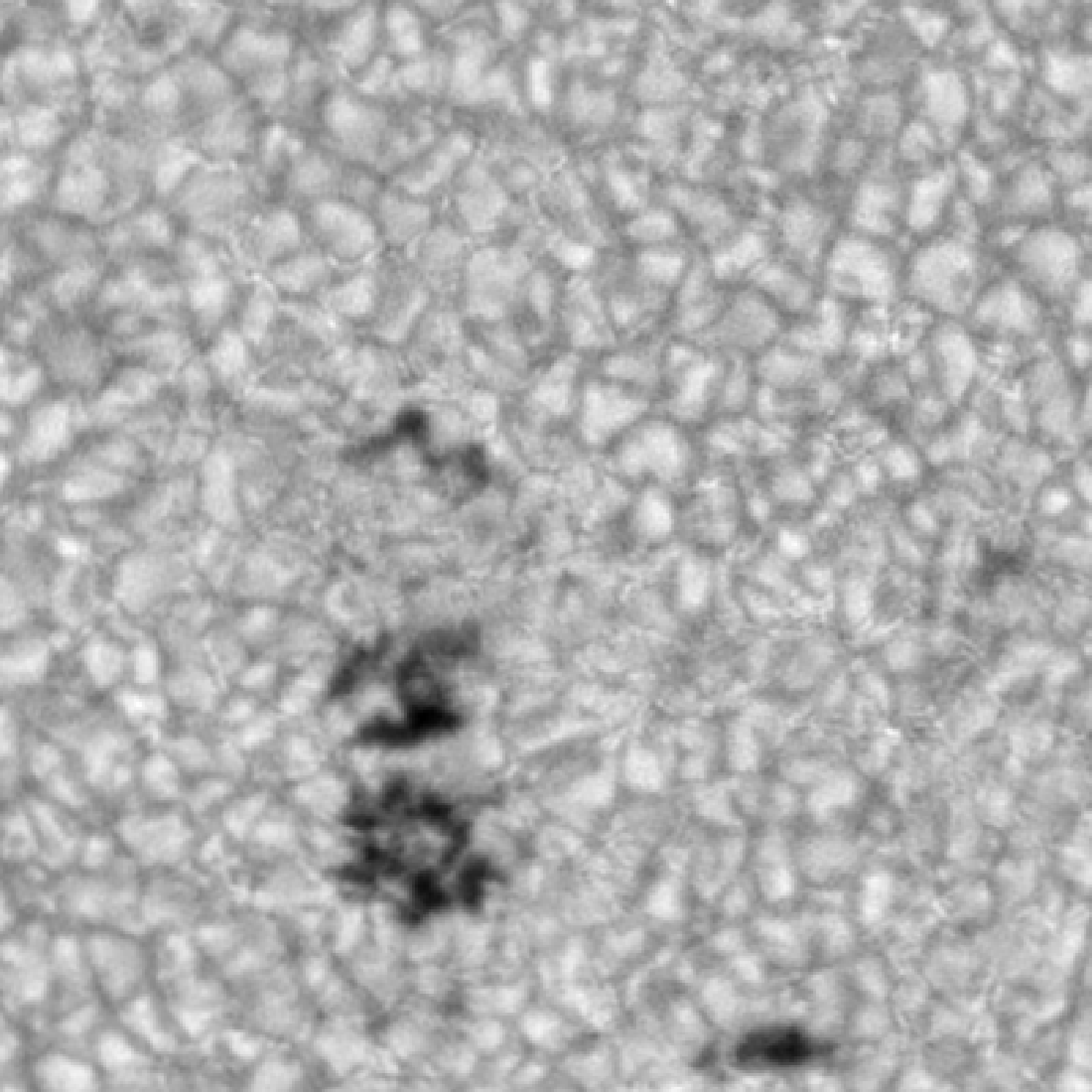}\end{minipage}

    \vspace{0.2em} 

    \begin{minipage}{0.195\textwidth}
        \centering
        \begin{overpic}[width=\textwidth]{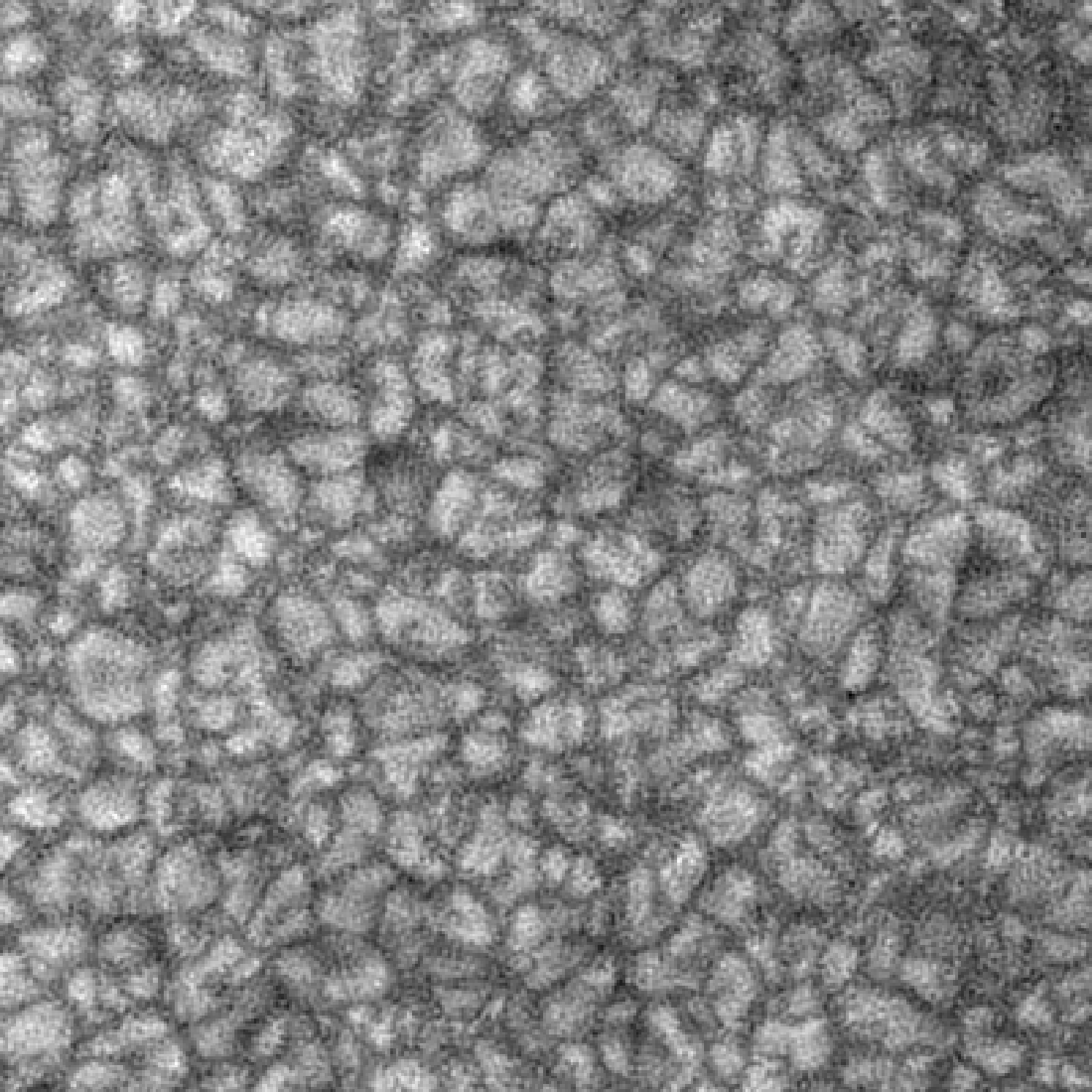}
            \put(30,5){\color{red}\large\textbf{10829\,\AA}}
        \end{overpic}
    \end{minipage}\hspace{0.2em}%
    \begin{minipage}{0.195\textwidth}\centering\includegraphics[width=\textwidth]{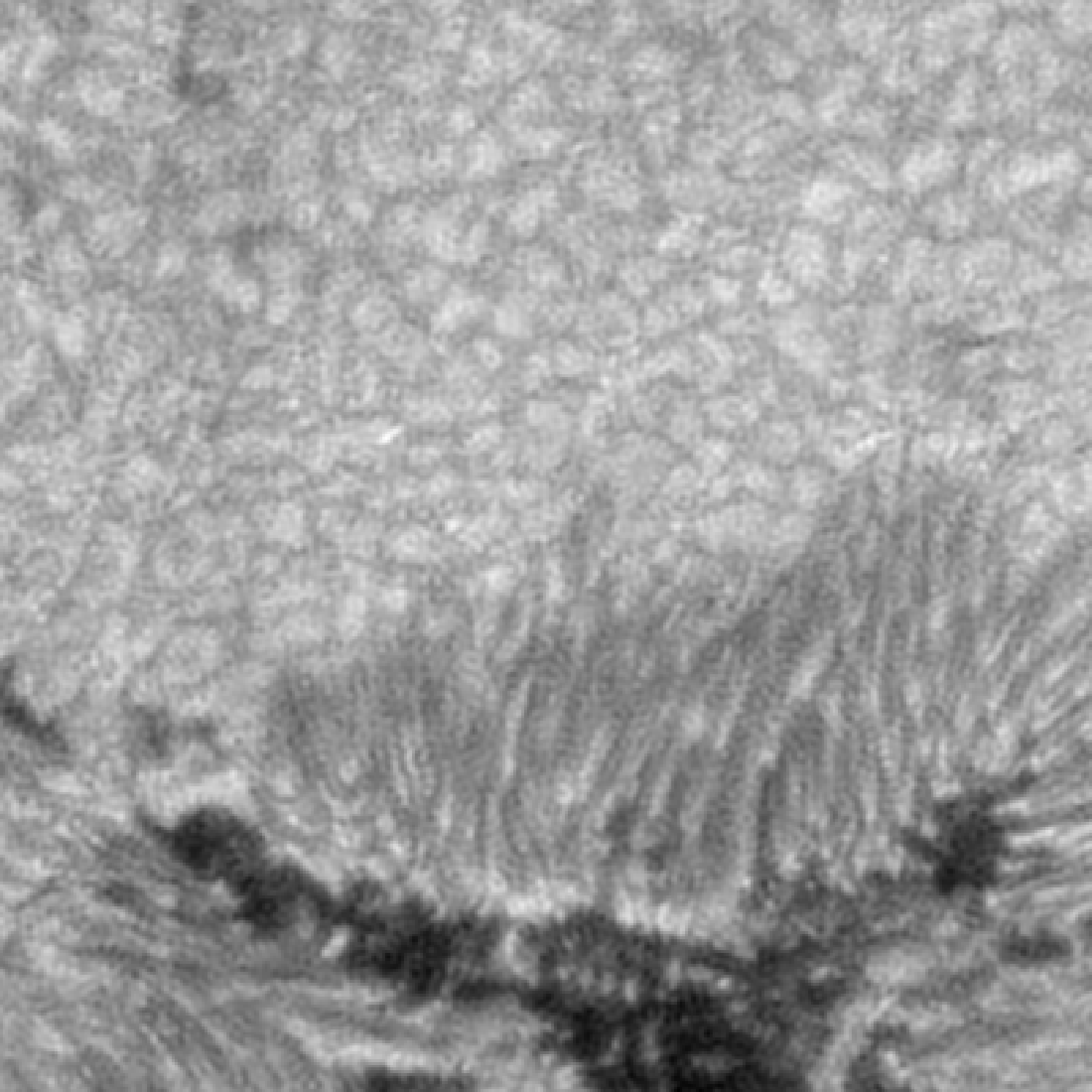}\end{minipage}\hspace{0.2em}%
    \begin{minipage}{0.195\textwidth}\centering\includegraphics[width=\textwidth]{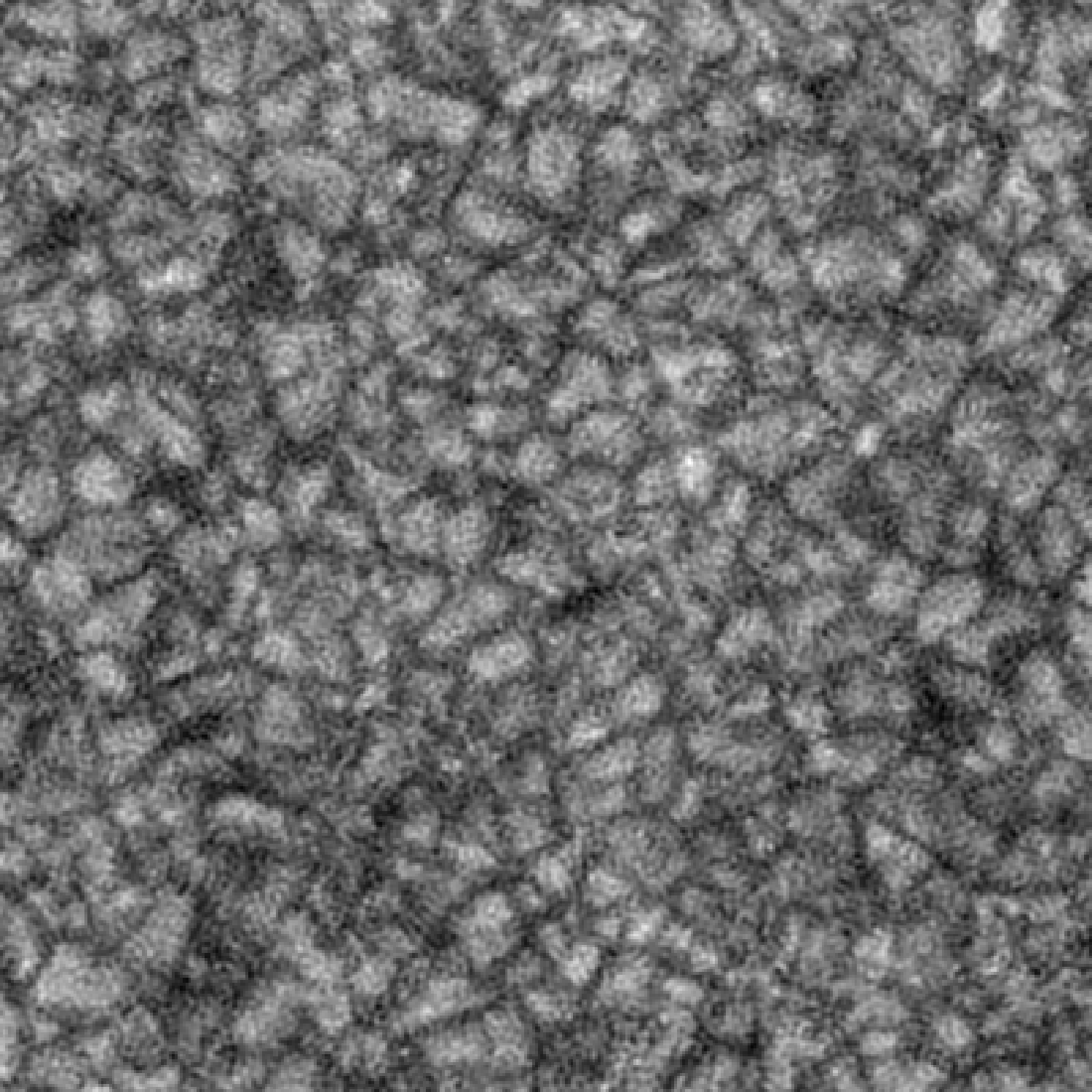}\end{minipage}\hspace{0.2em}%
    \begin{minipage}{0.195\textwidth}\centering\includegraphics[width=\textwidth]{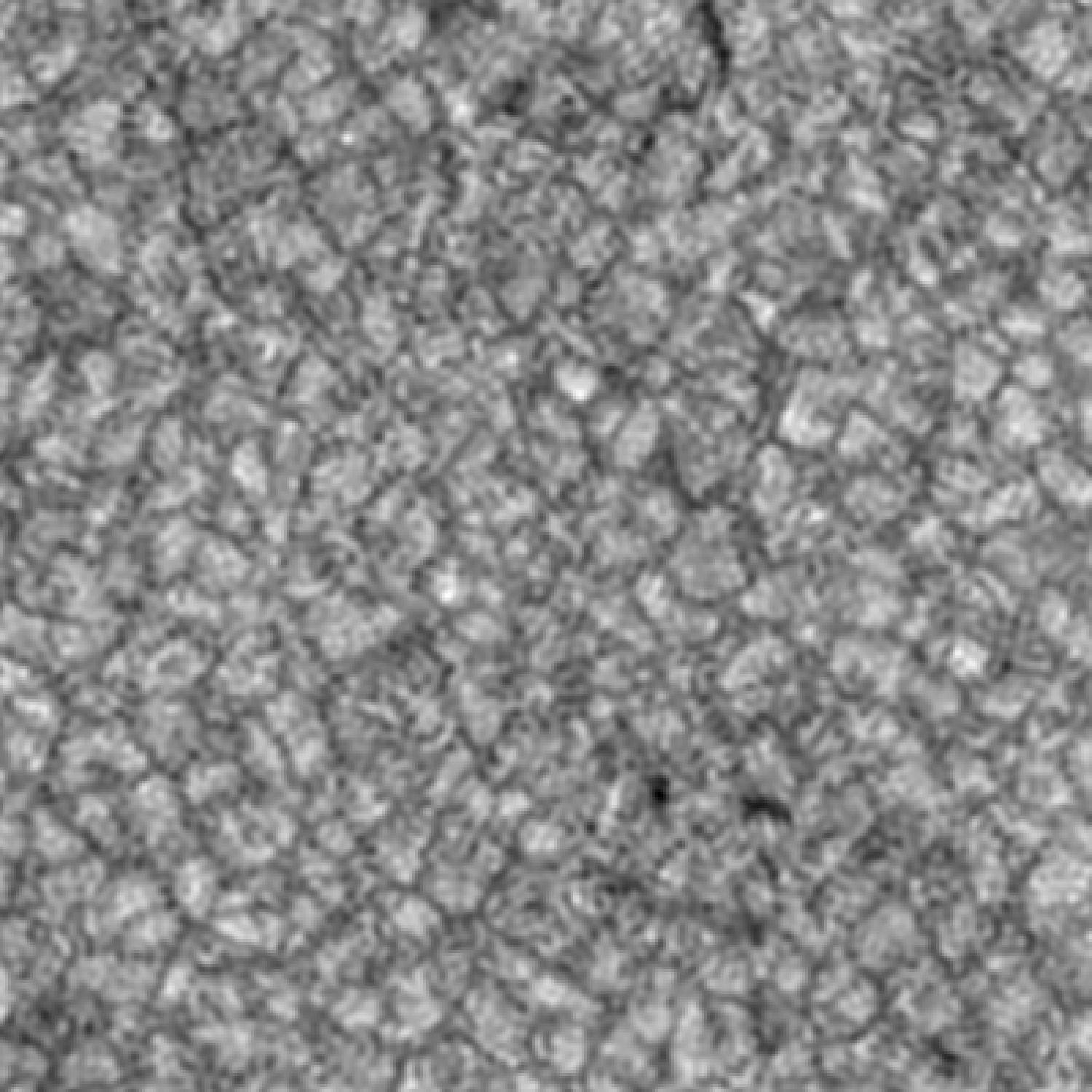}\end{minipage}\hspace{0.2em}%
    \begin{minipage}{0.195\textwidth}\centering\includegraphics[width=\textwidth]{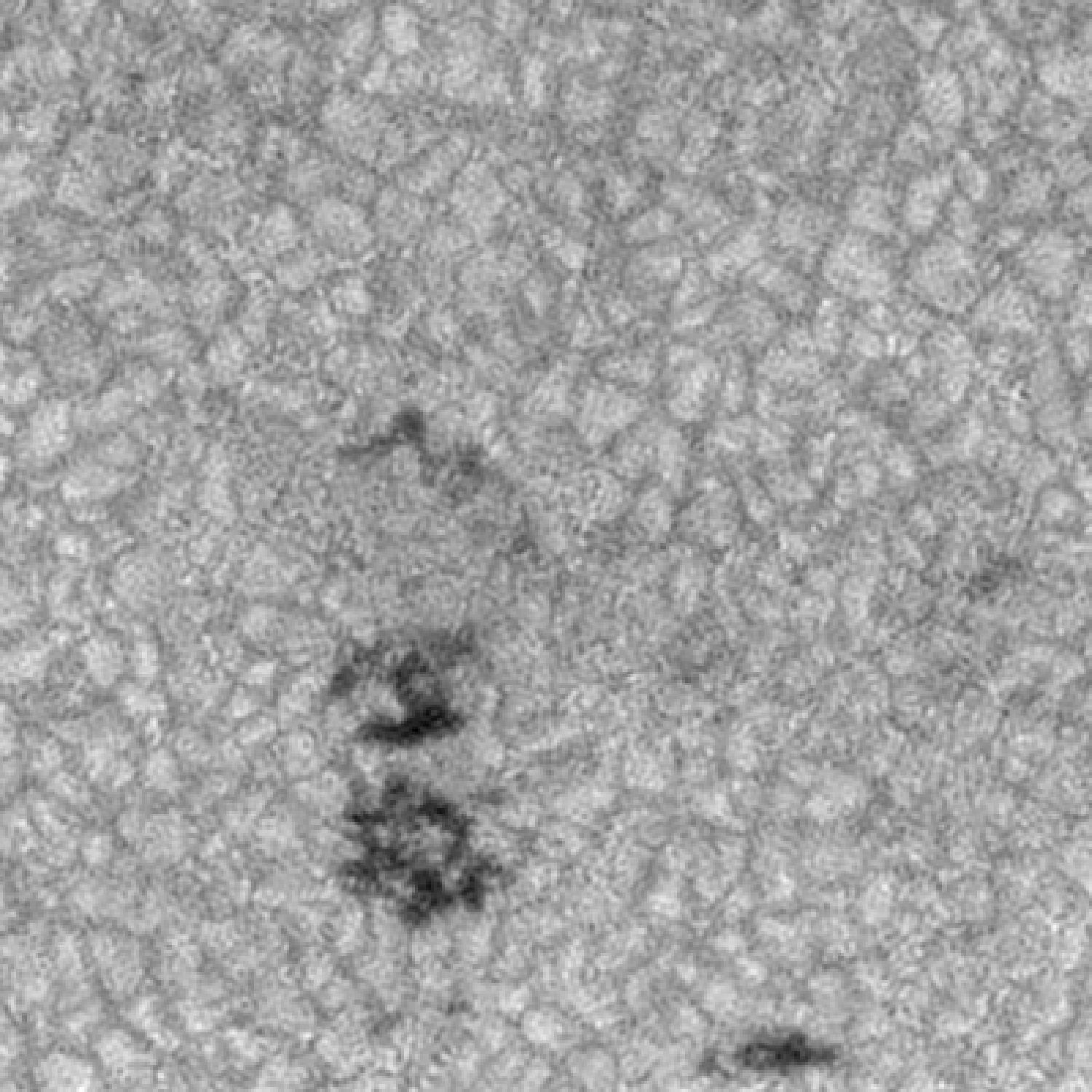}\end{minipage}

    \caption{Training data for Model A from NVST observations. From left to right: 2024-11-25, 2024-11-30, 2025-01-13, 2025-04-25, and 2025-05-12. The first row shows TiO images, and the second row shows 10829\,\AA\ images representing the photospheric component of \text{He~\textsc{i}}~10830~\text{\AA}.}
    \label{fig:2x5}
\end{figure*}

Model A is applied to the AR to establish a general mapping between $\text{TiO}$ images and the photospheric background signal in the $\text{10830\,\text{\AA}}$ band, using a training dataset that comprises TiO images and $\text{10829\,\text{\AA}}$ band observational data acquired over five distinct days. 
For typical AR features such as sunspots and filaments, the $\text{10830\,\text{\AA}}$ signals exhibit strong absorption with well-defined contours. However, the observational field of view still encompasses numerous magnetic loops, small jets, and faint chromospheric signals embedded within the strong photospheric background.
This model aims to extract these  chromospheric signals while achieving cross-dataset generalization. By using photospheric images and 10829\,\AA\ band data for training, the model generates photospheric reference images. Dividing the observed 10830\,\AA\ data by these references provides pixel-wise chromospheric information, enabling the prediction of photospheric reference signals and the precise extraction of chromospheric signals under different dates or observational conditions.
As shown in Figure~\ref{fig:2x5}, the \text{He~\textsc{i}}~10829~\text{\AA} band image reflects the photospheric characteristics of the observed region in terms of both structure and intensity distribution. 
This result provides the model with a high-quality reference label, allowing us to establish an accurate mapping relationship between the photospheric images in different bands.


Model B addresses the challenge posed by QS and weak absorption regions, where the \text{He~\textsc{i}}~10830~\text{\AA} chromospheric absorption is often too weak to be clearly distinguished under typical observational constraints. Model B treats the chromospheric signal as a subtle residual signal, extracting the difference between the raw observational data and the reconstructed background by learning the mapping relationship between two photospheric signals. This difference serves as an approximate representation of the chromospheric signal.
This approach is capable of efficiently and accurately separating out weak physical signals under low signal-to-noise ratio (SNR) conditions, demonstrating strong robustness and universality. During the training process, the model directly utilizes $\text{TiO}$ images (see $\text{B1}$ in Figure~\ref{fig:input_output}) to learn the photospheric equivalent signal of the $\text{10830\,\text{\AA}}$ band (see $\text{B2}$ in Figure~\ref{fig:input_output}), thereby achieving cross-band photospheric mapping and chromospheric residual signal extraction.


\subsection{Network}

Our method employs two dual-branch architectures based on Deep Convolutional Network models: $\text{Model A}$  and $\text{Model B}$ , which incorporates the Fourier Features Let Networks ($\text{FFLN}$) \citep{Cite23}.
Both models are fundamentally built upon the Deep Residual Network ($\text{ResNet}$) architecture \citep{Cite22}. The overall structure is composed of an initial convolutional layer, a subsequent series of Residual Blocks used for efficient feature extraction, and a final output layer. Specifically, Model B integrates the $\text{FFLN}$ module immediately following the input layer. The core objective of this architecture is to map the $\text{TiO}$ input to the target $\text{He I 10830\,\text{\AA}}$ corresponding photospheric signal.
In traditional ResNet architectures, components such as Batch Normalization (BN), Squeeze-and-Excitation (SE) blocks, and Dropout are modular options. In this study, the purpose of the dual-branch architecture is to use downsampling to account for details at different scales. The detailed configurations of Model A and Model B are in Table~\ref{tab:arch}.
Our $\text{ResNet}$ module implements a crucial enhancement to the classic residual learning paradigm, specifically by embedding the Channel Attention $\text{Squeeze-and-Excitation (SEBlock)}$ mechanism onto the main path of the residual connection. The $\text{SEBlock}$ dynamically learns and generates weighting coefficients for each feature channel, thereby significantly enhancing the model's representational capacity and its ability to focus on critical information.

\begin{deluxetable}{lccccc} 
\tablewidth{0pt}
\tablecaption{Comparison of Architectural Components for Model A and Model B \label{tab:arch}}
\tablehead{
\colhead{Model} & \colhead{SE Block} & \colhead{BN} & \colhead{Dropout} & \colhead{Branches} & \colhead{FFLN}
}
\startdata
Model A (AR) & Yes & Yes & YES & two & No \\
Model B (QS) & Yes & NO & No & two & Yes \\
\enddata

\end{deluxetable}




Model A adopts a dual-branch parallel structure to facilitate cross-scale feature learning. This design is intended to simultaneously learn granules and the finer intergranular lane structures, thereby balancing the extraction of local details and global contextual information. The architecture implements cross-scale fusion, where the High-Resolution Branch maintains the original input size ($400\times400$) to focus on capturing high-frequency details, such as granules and intergranular lanes. Meanwhile, the Mid-Resolution Branch downsamples the input to ($200\times200$) to process global contextual information and larger-scale granular structures, thus reducing computational complexity and increasing the receptive field.
Model A incorporates Dropout regularization after the ReLU activation within the residual blocks. This strategy effectively prevents the model from overfitting on lower-quality data, thereby enhancing its generalization capability. Feature Fusion: The feature maps from the two branches (each with $64$ channels) are upsampled and aligned, followed by Concatenation along the channel dimension (resulting in $128$ total channels), and finally mapped to the output through a $1\times1$ convolutional layer.


$\text{Model B}$ is an optimized variant of Model A, designed to overcome the Spectral Bias in deep learning and better learn high-frequency details. The design of Model B incorporates specific regularization and channel design choices. It does not utilize Dropout regularization and omits $\text{BN}$ layers within its Residual Blocks, while its channel count is set to double that of $\text{Model A}$. This distinct design stems from $\text{Model B}$'s specific training purpose: it uses observational data from only a single, particular day as its dataset, serving exclusively as a dedicated model for extracting the chromospheric signal from that day. The model does not need to address generalization across different dates or instruments under this special requirement, thus the removal of $\text{Dropout}$ and $\text{BN}$ is justified. Simultaneously, increasing the network capacity aims to maximize the model's ability to fit the specific day's training data, thereby preventing underfitting and ensuring the model captures all details within that day's data as precisely as possible.
FFLN Integration: This module maps the pixel's spatial coordinates ($\text{x}, \text{y}$) into high-dimensional sine and cosine features, which are then concatenated with the original image features along the channel dimension. This explicit spatial encoding mechanism significantly enhances the network's capacity to capture high-frequency information. Dimension Adjustment: Considering the additional feature dimensions and computational load introduced by $\text{FFLN}$, we adjusted $\text{Model B}$'s input image size to $200\times200$, and the downsampling target for the mid-resolution branch to $100\times100$.



\begin{figure*}[!htbp]
    \centering

    \begin{subfigure}[b]{0.4\textwidth}
        \centering
        \includegraphics[width=\textwidth]{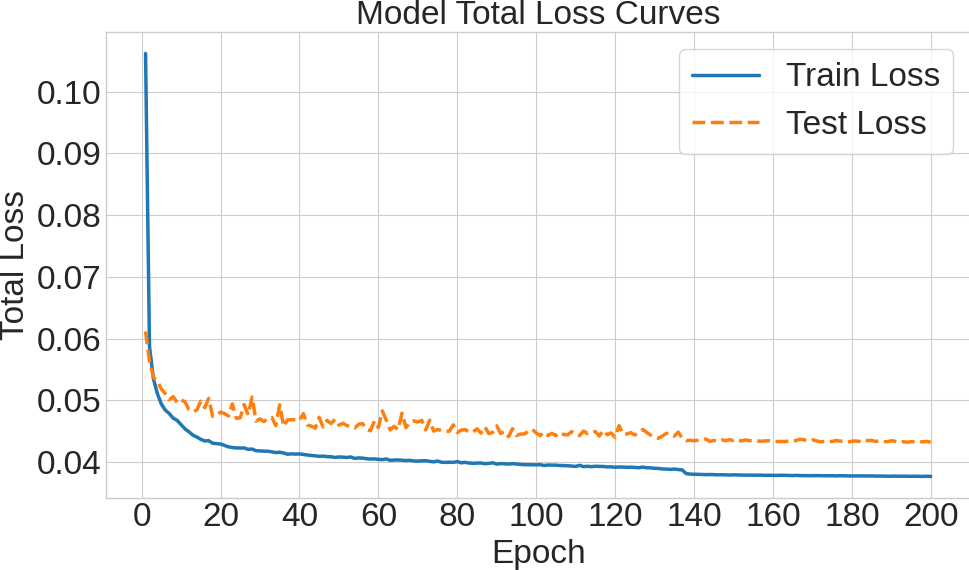}
        \caption{Model A Loss}
    \end{subfigure}
    \begin{subfigure}[b]{0.4\textwidth}
        \centering
        \includegraphics[width=\textwidth]{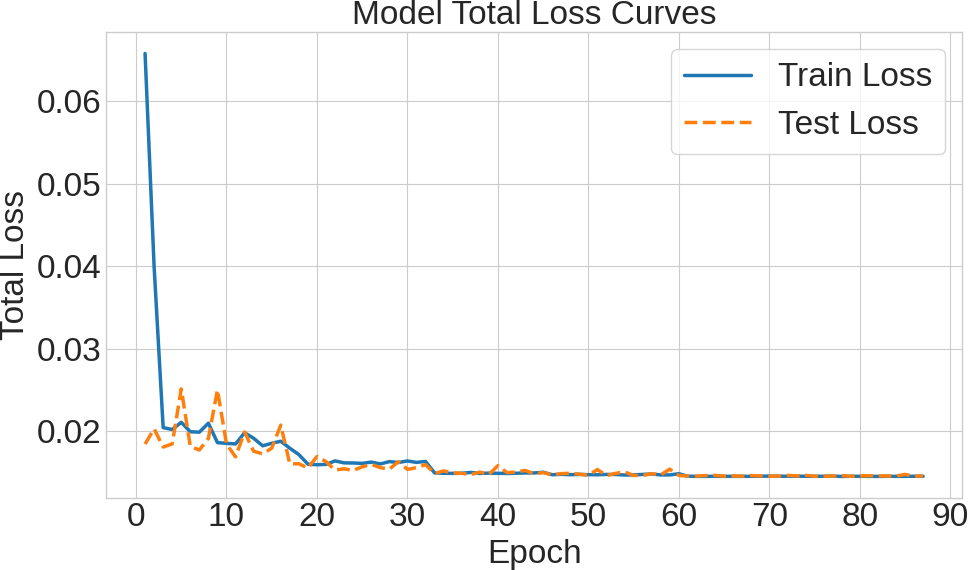}
        \caption{Model B Loss}
    \end{subfigure}

    \caption{\textbf{Comparison of Loss for Model A and Model B.}}
    \label{fig:models_combined}
\end{figure*}

As shown in Figure~\ref{fig:models_combined}, the total loss function is composed of an $\text{L}_1$ loss and a weighted gradient loss ($0.3 \times \mathcal{L}_{\text{grad}}$). The gradient loss enforces the model to learn and preserve high-frequency information such as edge details and texture variations in images by jointly optimizing the horizontal and vertical gradient discrepancies between the predicted and target outputs.
The models were implemented using the PyTorch framework and optimized using the Adam optimizer, and cross-validation was not adopted in this study.
In practice, the $\text{L}_1$ term accounts for approximately half of the total loss. 
$\text{Model A}$ started training with a learning rate of $10^{-3}$ and converged to $10^{-4}$. We attempted to further reduce the learning rate, but the results showed no significant change in performance, thereby validating the effectiveness of the current setting. $\text{Model B}$ was initialized with a learning rate of $10^{-4}$ and converged to $10^{-6}$. This small final learning rate is appropriate and indicates that the training process achieved good convergence.
Regarding performance metrics, the Mean Absolute Error ($\text{MAE}$) of $\text{Model A}$ is approximately $\text{0.015}$. Since $\text{Model B}$ is a dedicated model for high-precision fitting of single-day data, its $\text{MAE}$ is even lower, actually falling below $\text{0.01}$ at approximately $\text{0.007}$. Our models were trained for sufficient epochs so that both the training loss and the validation loss gradually approach a nearly stable minimum. This demonstrates that the models not only achieved good training outputs but that $\text{Model A}$ also exhibited excellent generalization capability.



\begin{figure*}[htbp]
    \centering
    \includegraphics[width=0.98\textwidth]{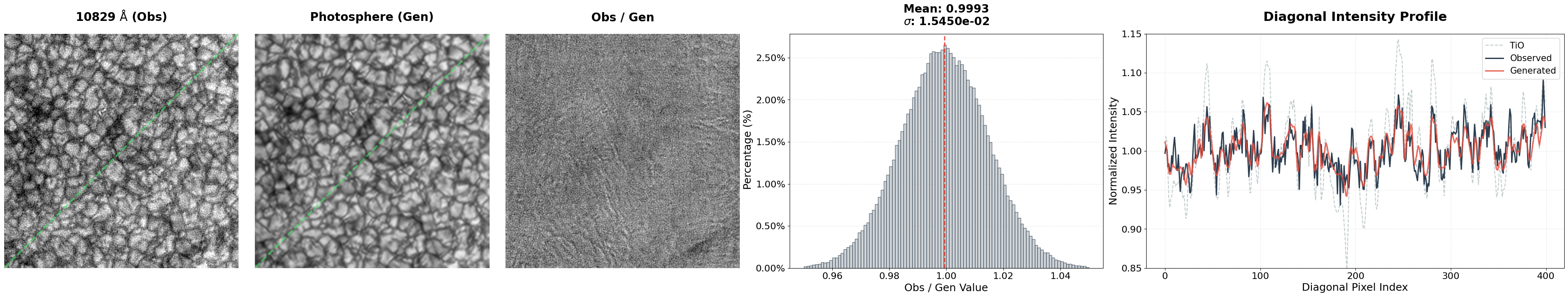}
    \includegraphics[width=0.98\textwidth]{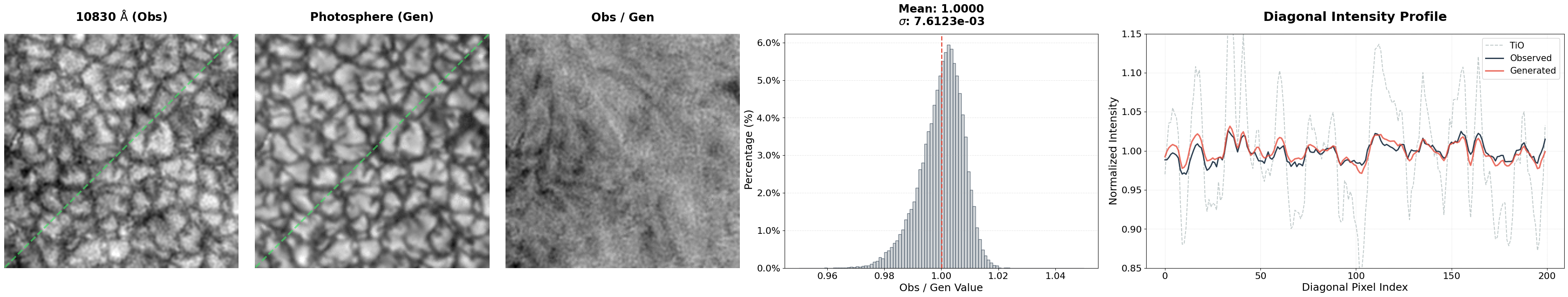}
    
    \caption{Testing cases of our proposed models: The top row displays results from Model A for active regions, and the bottom row shows results from Model B for the quiet Sun. 
    The columns from left to right represent: 
    (1) the observed intensity used as the training label (Obs); 
    (2) the predicted 10830 \text{\AA} photospheric signal generated by the model (Gen); 
    (3) the ratio map indicating relative deviations (Obs/Gen); 
    (4) the distribution of the Obs/Gen values representing the model's prediction accuracy; and 
    (5) the intensity values of pixels sampled along the diagonal from the bottom-left to the top-right of the images.}
    \label{fig:test_cases}
\end{figure*}

As illustrated in Figure~\ref{fig:test_cases}, the relative errors ($\sigma$) for mapping the 10830~{\AA} photospheric information are approximately 0.015 for Model~A and 0.007 for Model~B. The magnitude of this statistical error is primarily limited by residual chromospheric signals within the observed labels (Obs). Since the model-generated output (Gen) represents a pure photospheric background, any residual chromospheric absorption features in the labels result in localized increases in residuals (e.g., reaching $\sim$0.025 for Model~A and $\sim$0.01 for Model~B). Such fluctuations do not signify a decrease in accuracy; rather, they reflect the model's effective decoupling of non-photospheric interference. 
As observed from the diagonal intensity profiles, TiO and Obs exhibit similar variation trends in brightness, indicating they share the same underlying morphological structures. However, it is evident that the relationship between them is not a simple linear mapping. Therefore, non-linear model training is of critical importance for such tasks. The variation trends and absolute magnitudes of the Gen curves match the Obs profiles extremely closely, demonstrating that the model has successfully captured the complex non-linear mapping from TiO to 10830~{\AA} and achieved high precision.



\subsection{Chromospheric Signal Separation}

\begin{figure*}[!htbp]

    \centering

        \begin{minipage}{0.2\textwidth}
            \centering
            \begin{overpic}[width=\textwidth]{AR/input_tensor_0081.png}
                \put(2,5){\color{red}\textbf{TiO (Model A input)}}
            \end{overpic}
        \end{minipage}\hspace{0.3em}%
        \begin{minipage}{0.2\textwidth}
            \centering
            \begin{overpic}[width=\textwidth]{AR/output_tensor_0081.png}
                \put(10,5){\color{red}\textbf{He I (observation)}}
            \end{overpic}
        \end{minipage}\hspace{0.3em}%
        \begin{minipage}{0.2\textwidth}
            \centering
            \begin{overpic}[width=\textwidth]{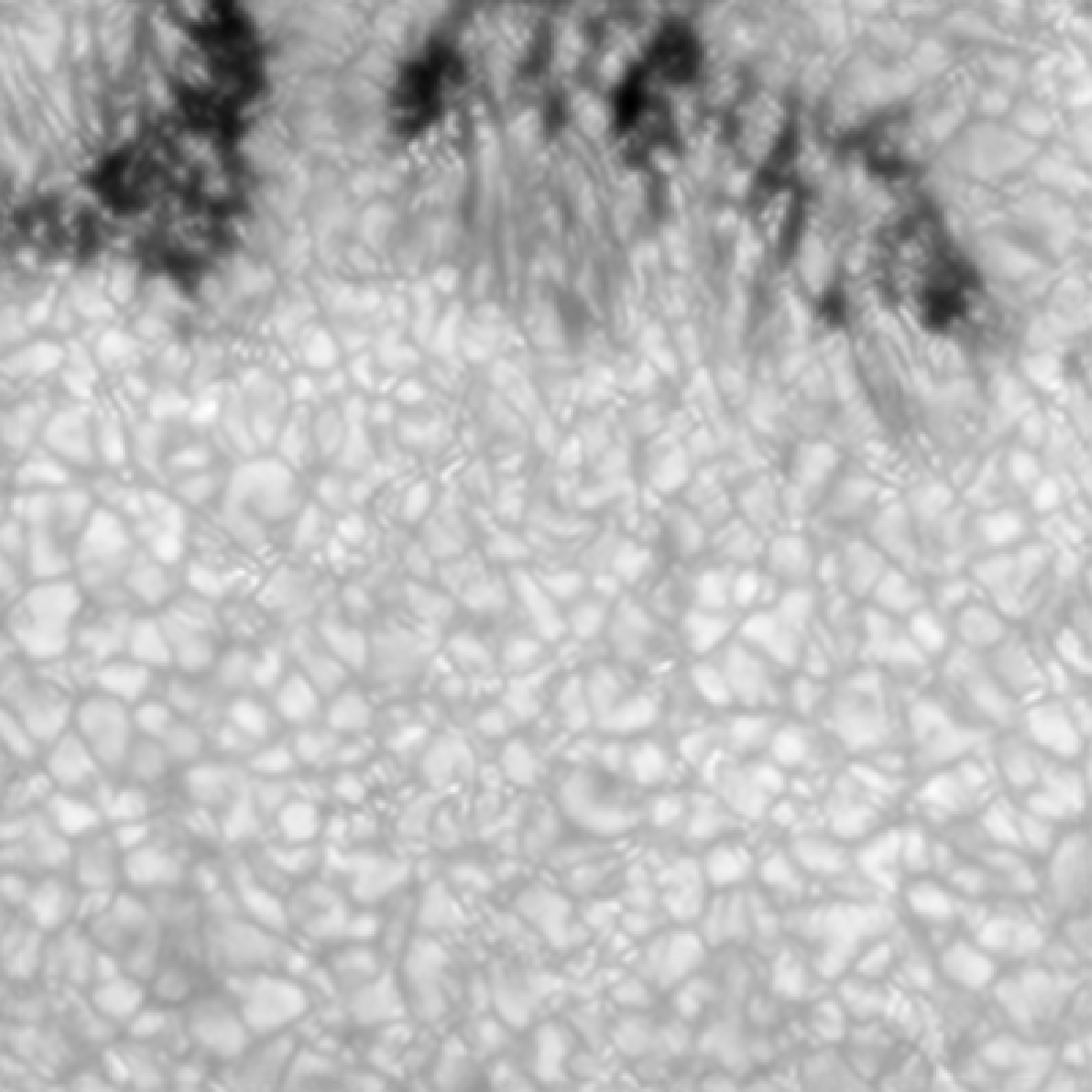}
                \put(20,5){\color{red}\textbf{Photospheric}}
            \end{overpic}
        \end{minipage}\hspace{0.3em}%
        \begin{minipage}{0.2\textwidth}
            \centering
            \begin{overpic}[width=\textwidth]{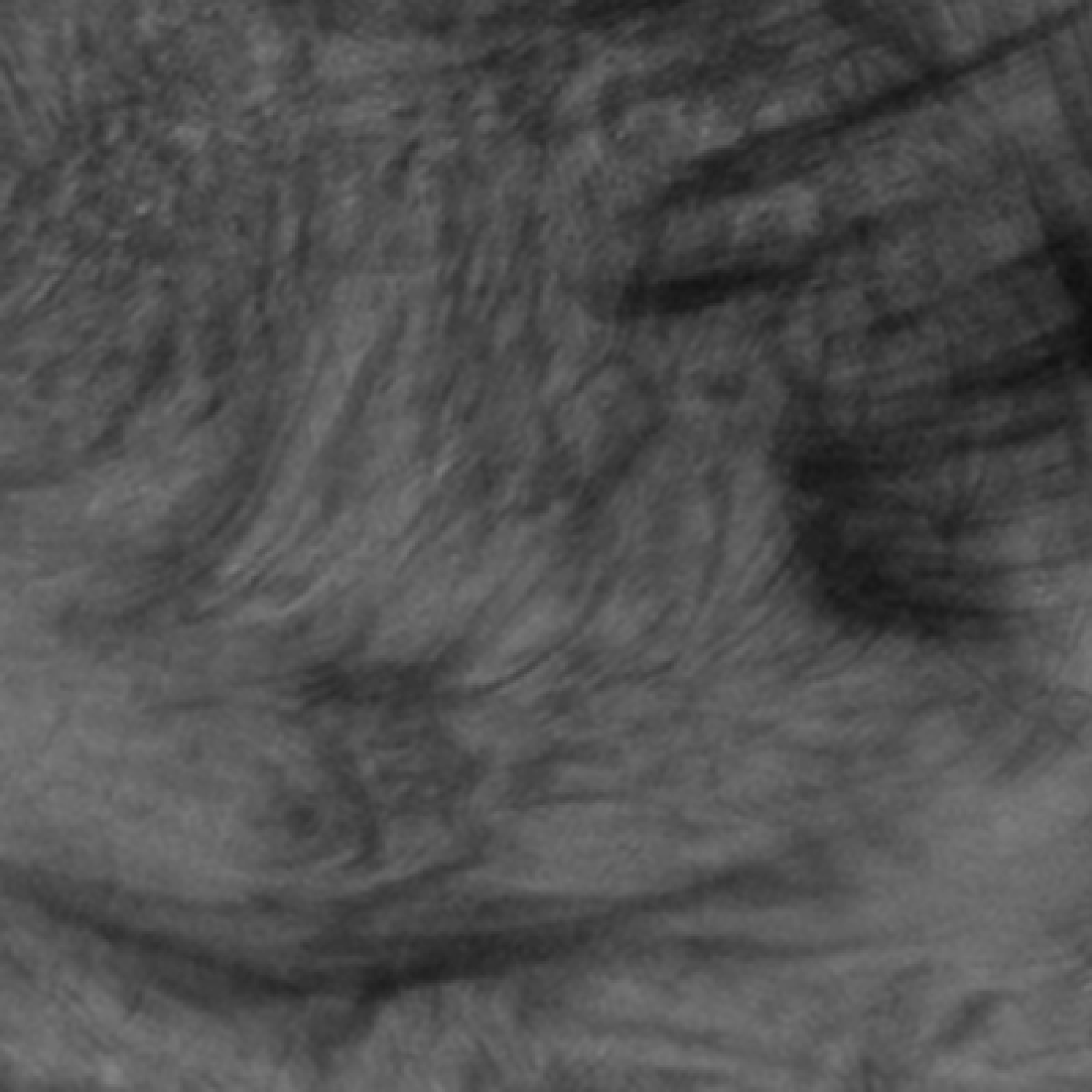}
                \put(20,5){\color{red}\textbf{Chromospheric}}
            \end{overpic}
        \end{minipage}
    \vspace{3pt} 


        \begin{minipage}{0.2\textwidth}
            \centering
            \begin{overpic}[width=\textwidth]{QS/Tio_20241119_053046_cut.png}
                \put(2,5){\color{red}\textbf{TiO (Model B input)}}
            \end{overpic}
        \end{minipage}\hspace{0.3em}%
        \begin{minipage}{0.2\textwidth}
            \centering
            \begin{overpic}[width=\textwidth]{QS/HeI_20241119_053045_cut.png}
                \put(10,5){\color{red}\textbf{He I (observation)}}
            \end{overpic}
        \end{minipage}\hspace{0.3em}%
        \begin{minipage}{0.2\textwidth}
            \centering
            \begin{overpic}[width=\textwidth]{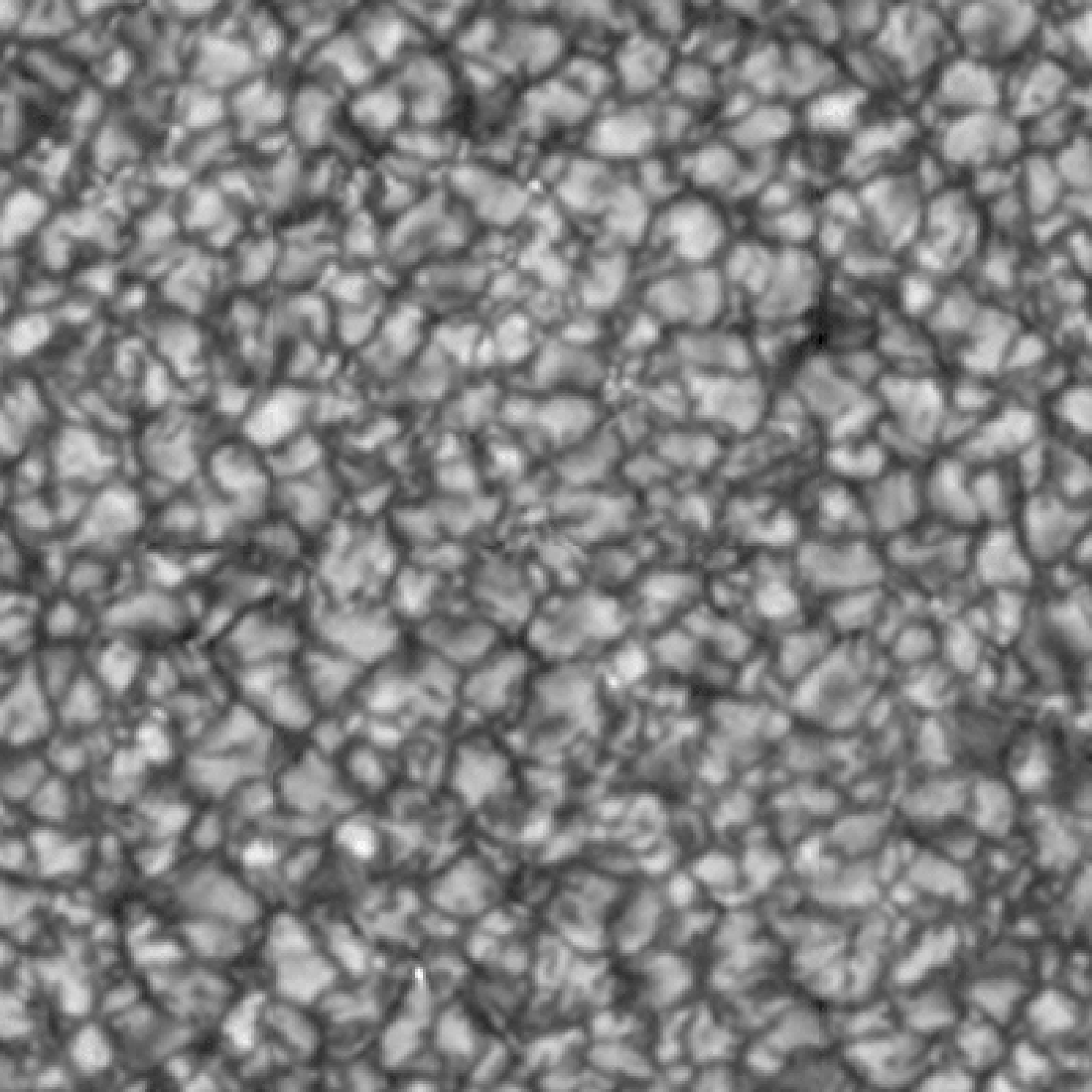}
                \put(20,5){\color{red}\textbf{Photospheric}}
            \end{overpic}
        \end{minipage}\hspace{0.3em}%
        \begin{minipage}{0.2\textwidth}
            \centering
            \begin{overpic}[width=\textwidth]{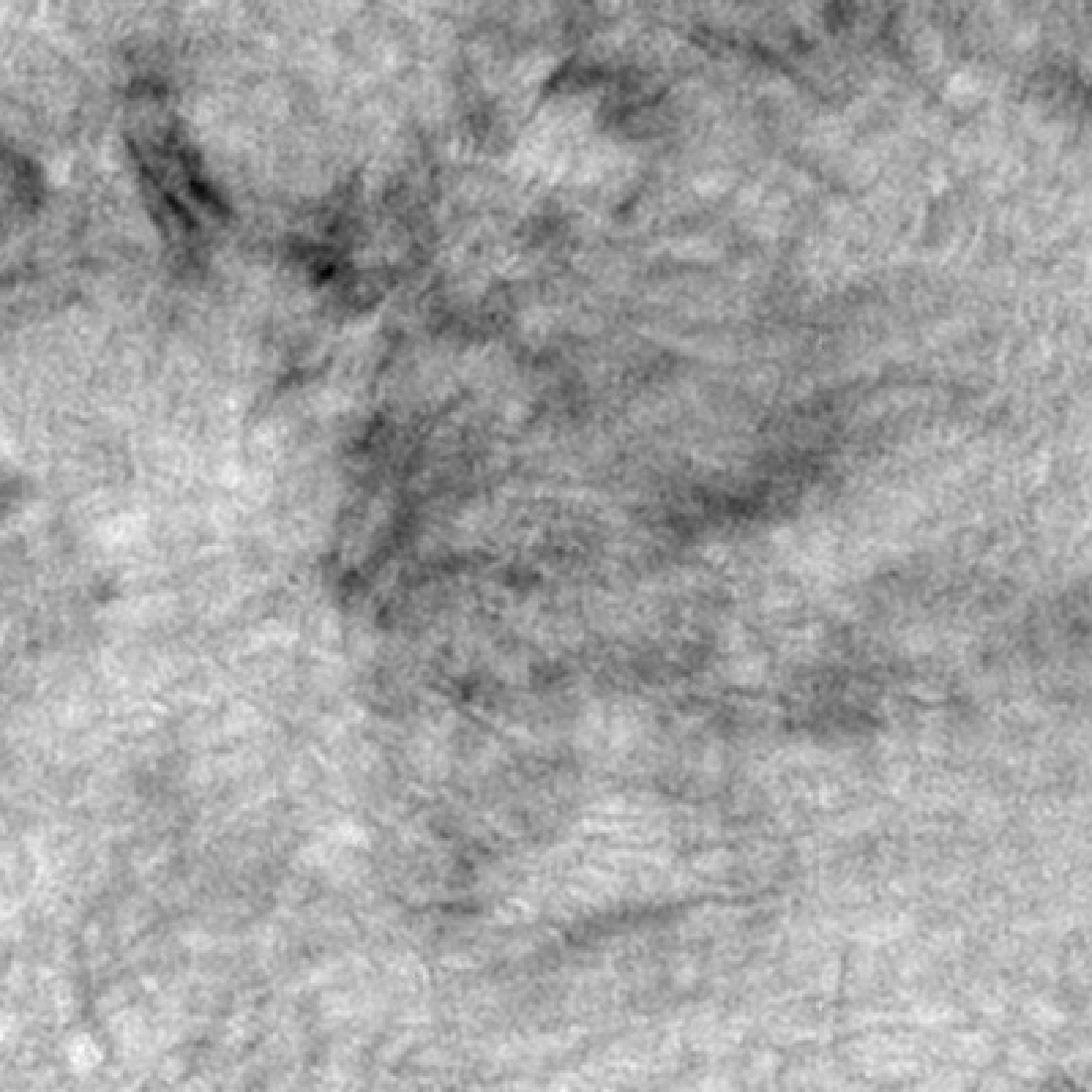}
                \put(20,5){\color{red}\textbf{Chromospheric}}
            \end{overpic}
        \end{minipage}
    \caption{Comparison of photospheric and chromospheric components.
    The first row shows the active region; 
    the second row shows the quiet region. 
    Columns 1–2 display the original TiO and \text{He~\textsc{i}}~10830~\text{\AA} observations; 
    columns 3–4 present the generated photospheric reference and extracted chromospheric absorption maps.
    All images have a spatial size of $400\times400$ pixels.
    Animations of the aligned sequences are available in the HTML version for the AR ($300 \times 300$ pixel, 3 s) and QS ($400 \times 400$ pixel, 4 s) regions. These animations cover $\sim$40–45 minutes of solar evolution, highlighting the dynamic chromospheric fibrils (Column 4) isolated from the composite signals (Column 2)}
    
    \label{fig:prediction_separation}
    
\end{figure*}
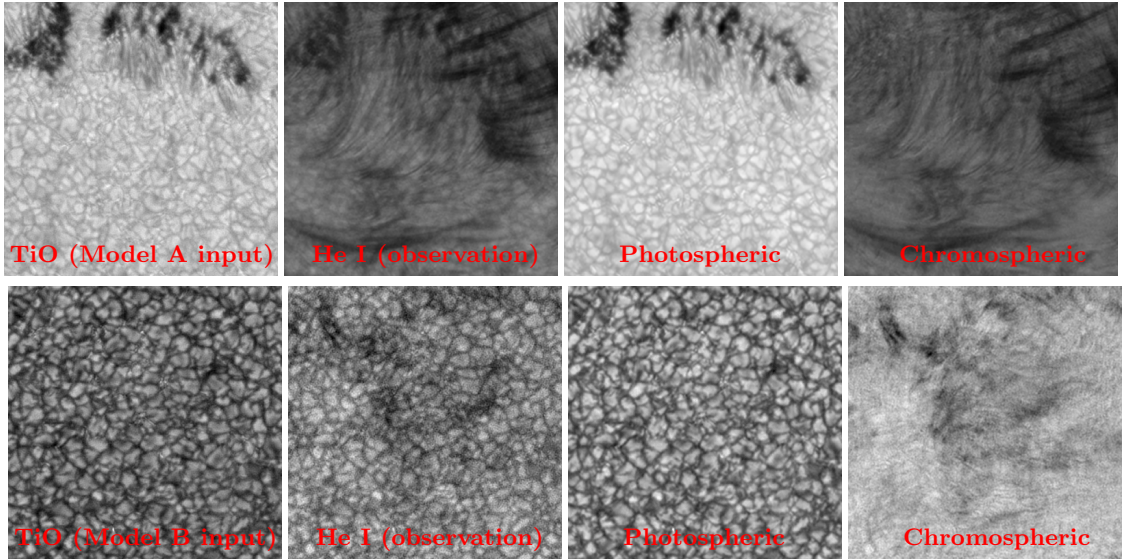

Based on the trained and validated deep learning model, the $\text{TiO}$ data ( Figure~\ref{fig:prediction_separation}, Column 1) is utilized to predict and generate the corresponding photospheric image in the \text{He~\textsc{i}}~10830~\text{\AA} band (Figure~\ref{fig:prediction_separation}, Column 3) without requiring a target label. This prediction step is crucial as it establishes the necessary photospheric reference for subsequent calculations. Subsequently, the final chromospheric signal (Figure~\ref{fig:prediction_separation}, Column 4) is separated and extracted by applying a simple exponential absorption model to the observed \text{He~\textsc{i}}~10830~\text{\AA} data (Figure~\ref{fig:prediction_separation}, Column 2) and the generated photospheric image. Critically, this physical model provides the genuine physical basis for decoupling the chromospheric signal from the continuum. In the following section, we analyze the chromospheric activity within a quiet region characterized by relatively weak chromospheric emission.


\section{Results}\label{sec:Results}

\begin{figure*}[htb!] 
    \centering 
    
    \includegraphics[width=0.9\linewidth]{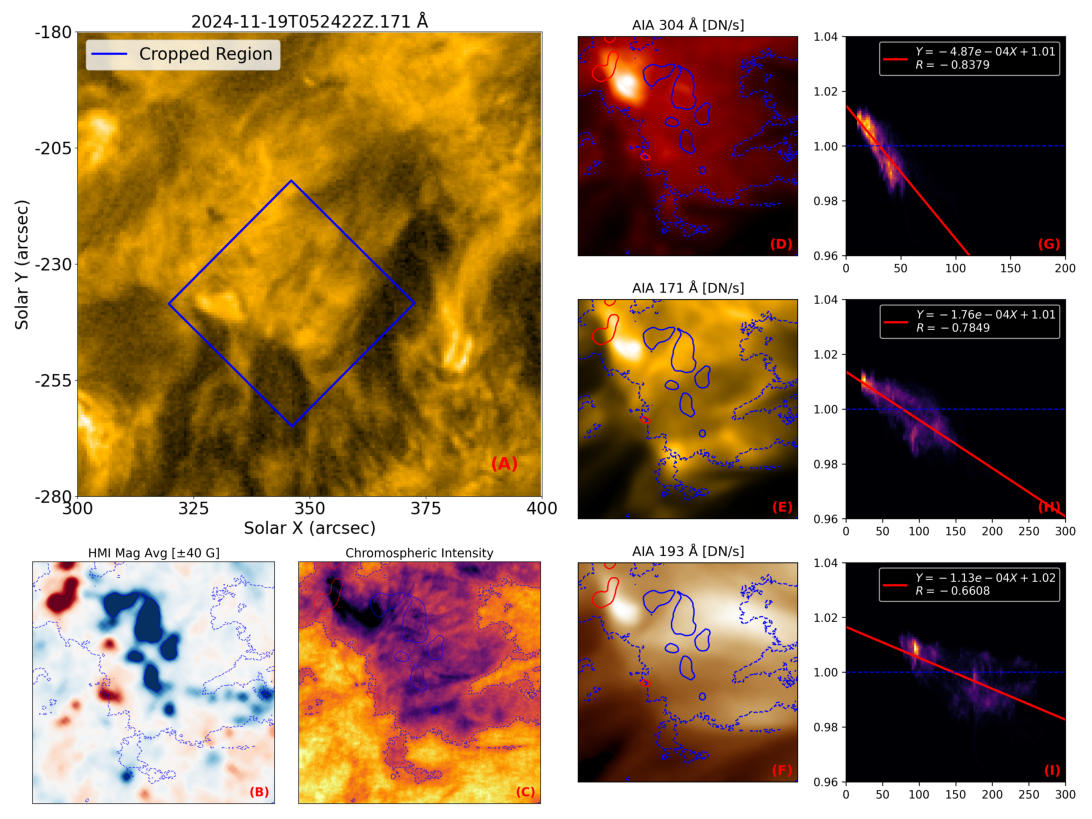}
    \caption{Multi-wavelength observations and quantitative analysis of the quiet-Sun region. The images used in this figure are 45-minute time-averaged data.
    (A) $\text{SDO/AIA}$ $\text{171\,\text{\AA}}$ image, with the blue box indicating the area of interest for this study;
    (B) $\text{HMI}$ magnetogram;
    (C) Chromospheric structure separated from the \text{He~\textsc{i}}~10830~\text{\AA} line;
    (D)-(F) $\text{304\,\text{\AA}}$, $\text{171\,\text{\AA}}$, and $\text{193\,\text{\AA}}$ observations within the study region;
    (G)-(I) Scatter plots and linear fitting relationships, where the $\text{x}$-axis represents the intensity of the three $\text{AIA}$ bands ($\text{AIA 304\,\text{\AA}}$, $\text{171\,\text{\AA}}$, $\text{193\,\text{\AA}}$), and the $\text{y}$-axis represents the \text{He~\textsc{i}}~10830~\text{\AA} chromospheric signal.
    Legend Details: In subfigures (B)-(F), the blue dashed line represents the contour of the time-averaged mean intensity of the \text{He~\textsc{i}}~10830~\text{\AA} data, with the area inside the dashed line denoting the \text{He~\textsc{i}}~10830~\text{\AA} absorption region; the red solid line and the blue solid line denote the magnetic field contours of positive $\text{40\,G}$ and negative $\text{40\,G}$, respectively.
    }
    \label{fig:area}
\end{figure*}

On November 19, 2024, the observational data analyzed originated from a solar quiet region observation conducted between 05:25–06:10 (UT). This joint dataset provides multi-layered information, spanning from the photosphere to the corona, and comprises $\text{NVST}$ high-resolution imaging data ($\text{TiO}$ and \text{He~\textsc{i}}~10830~\text{\AA}) and $\text{SDO}$ data ($\text{AIA}$ at $171\,\text{\AA}$, $193\,\text{\AA}$, $304\,\text{\AA}$, and $\text{HMI}$'s $I_{\text{c}}$ and $I_{\text{m}}$).
Before analysis, the $\text{NVST}$ data were normalized using the average intensity of the observation sequence. The $\text{SDO}$ data (AIA and HMI), having been calibrated to represent physical quantities such as temperature, did not require further normalization.
The observational region for this study is contained within the blue box in Fig.~\ref{fig:area}A. 
The alignment process utilized a combination of the $\text{SIFT}$ method for robust feature matching and $\text{Optical Flow}$ for fine-scale image registration.
For temporal synchronization, $I_{\text{m}}$ were used as the reference, and other bands were matched based on the nearest-neighbor principle. 
For spatial alignment, the $10830\,\text{\AA}$ image was used as the reference frame: first, TiO was aligned to $10830\,\text{\AA}$ using SIFT and Optical Flow; subsequently, IDL procedures were used to achieve self-alignment of the SDO data, and SIFT+Optical Flow was applied again to align $I_c$ to TiO. Based on the resulting transformation matrix, HMI ($I_{\text{c}}$ and $I_{\text{m}}$) and AIA ($171\,\text{\AA}$, $193\,\text{\AA}$, $304\,\text{\AA}$) data were then aligned to \text{He~\textsc{i}}~10830~\text{\AA}. Bilinear interpolation was used during the alignment process to ensure pixel correspondence without changing the resolution of the data.



\subsection{Correlation between \texorpdfstring{$10830\text{ \AA}$}{10830 Å} and EUV Intensity}\label{subsec:Results1}


Our observations originate from the region delimited by the blue bounding box in the $\text{QS}$ area shown in $\text{Figure~\ref{fig:area}(A)}$. 
Fig.~\ref{fig:area}~(B)-(F) display the corresponding 45-minute averaged HMI magnetic field, \text{He~\textsc{i}}~10830~\text{\AA} chromosphere, $\text{AIA}$ $304\text{ \AA}$, $171\text{ \AA}$, and $193\text{ \AA}$ data for this observed region. 
The area enclosed by the blue dashed line represents the \text{He~\textsc{i}}~10830~\text{\AA} absorption region, with the threshold set to the average value of Figure~\ref{fig:area}C (value = 1), while the red and blue solid lines denote the boundaries of the $\pm 40\,\mathrm{G}$ magnetic fields, respectively.
We obtained a \text{He~\textsc{i}}~10830~\text{\AA} chromospheric signal (Figure~\ref{fig:area}~(C)) using a deep learning method and performed a comparative analysis with the intensity enhancement of three $\text{AIA}$ $\text{EUV}$ bands (Figure~\ref{fig:area}~(D)-(F)).
We found that, in the QS, the intensity of the \text{He~\textsc{i}}~10830~\text{\AA} chromospheric signal exhibits a significant negative spatial correlation with the $\text{EUV}$ band intensity. 
Furthermore, the chromospheric absorption regions and coronal brightening regions spatially overlap.
This strong negative correlation and spatial overlap strongly support a tight vertical coupling between energy release and response processes in the photosphere and corona.
On the 45-minute time-averaged data, the correlation coefficient $R$ ranges from $-0.84$ to $-0.66$ (Figure~\ref{fig:area}~(G)–(I)).
The \text{He~\textsc{i}}~10830~\text{\AA} signal and the $\text{AIA}\,\text{304\,\text{\AA}}$ emission show the strongest anti-correlation ($R \approx -0.84$). The anti-correlation weakens with increasing characteristic temperature, reaching $R \approx -0.78$ in the  $\text{AIA}\,\text{171\,\text{\AA}}$channel and $R \approx -0.66$ in the  $\text{AIA}\,\text{193\,\text{\AA}}$ channel.
This suggests that the correlation between \text{He~\textsc{i}}~10830~\text{\AA} absorption and EUV brightening decreases with temperature and formation height, and that in quiet regions, it is more strongly correlated with EUV radiation from the lower atmospheric layers.


\begin{figure*}[htb!]
    \centering
    \includegraphics[width=0.9\linewidth]{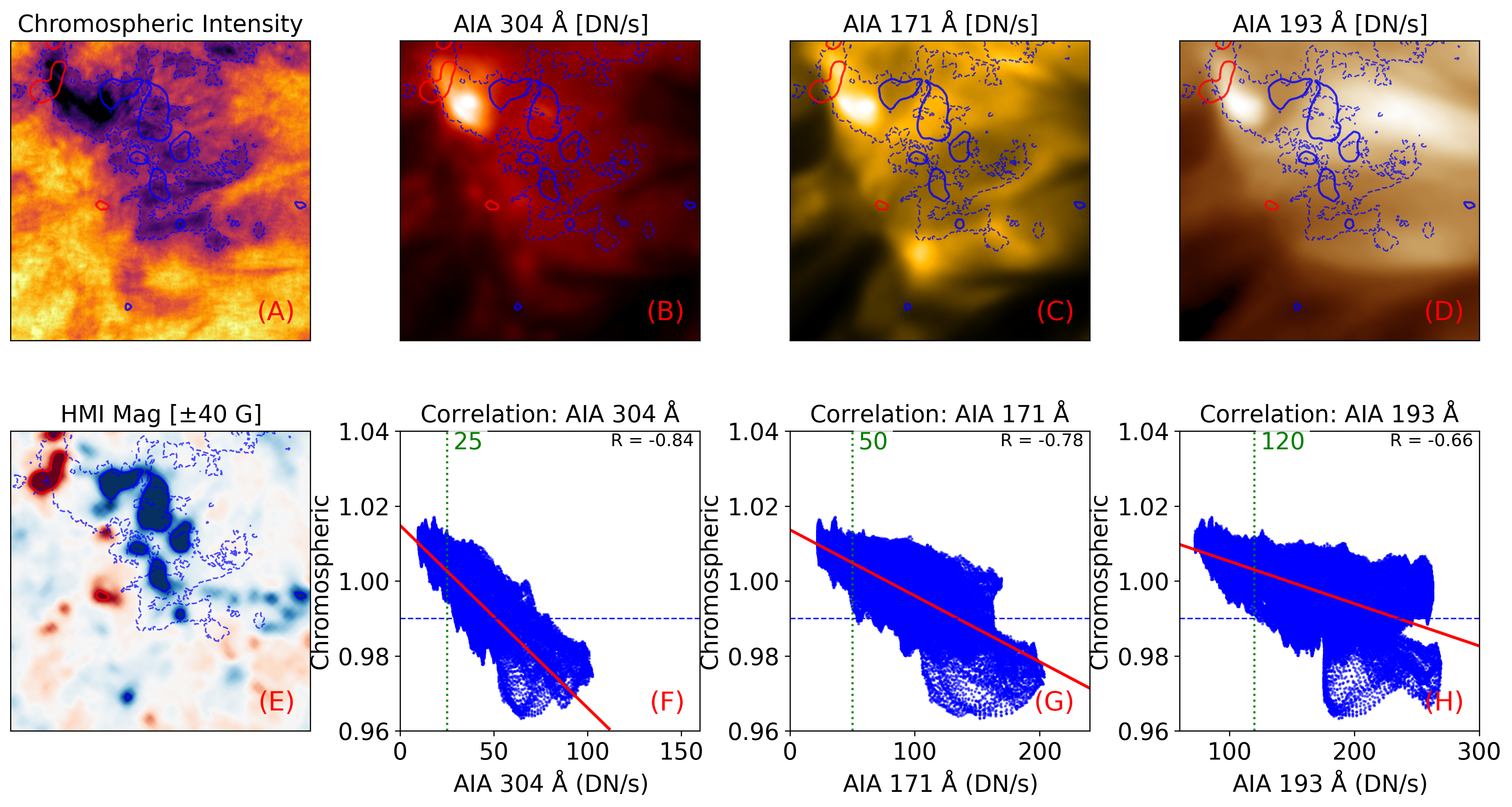}
    
    \caption{
        The images used in this figure are 45-minute time-averaged data.
        (A) The \text{He~\textsc{i}}~10830~\text{\AA} chromospheric signal;
        (B)-(D) $\text{AIA}$ band images at $\text{304\,\text{\AA}}$, $\text{171\,\text{\AA}}$, and $\text{193\,\text{\AA}}$, respectively;
        (E) The $\text{HMI}$ magnetogram.
        (F)-(H) Scatter plots and linear relationships between the three $\text{EUV}$ bands and the \text{He~\textsc{i}}~10830~\text{\AA} chromospheric signal.
        Legend Details: In panels (A)-(E), the blue dashed line outlines the strong absorption region, defined as the $\text{99\%}$ percentile of the time-averaged \text{He~\textsc{i}}~10830~\text{\AA} chromospheric signal;the red and blue solid lines represent the magnetic field contours of $\pm 40\,\mathrm{G}$;
        In panels (F)-(H), data points below the blue dashed line correspond to the strong absorption region, and the green dashed line shows the minimum $\text{EUV}$ brightness within this region.}
    \label{fig:about_Im}
\end{figure*}

\subsection{Thresholded EUV Response in \text{He~\textsc{i}}~10830~\text{\AA} Strong Absorption Regions}
\label{subsec:Results2}

The strong absorption regions of the chromospheric \text{He~\textsc{i}}~10830~\text{\AA}  line do not always coincide with the brightest $\text{EUV}$ regions. However, the $\text{EUV}$ intensity within these regions can still reach a threshold. 
These strong absorption regions are defined as the 99\% percentile of the time-averaged \text{He~\textsc{i}}~10830~\text{\AA} signal (value = 0.99), as indicated by the blue dashed box in Figure~\ref{fig:about_Im}~(A)-(E).
Within the strong absorption region, the $\text{EUV}$ bands generally exceed thresholds of $\sim 25\,\mathrm{DN/s}$ for AIA 304~\text{\AA}, $\sim 50\,\mathrm{DN/s}$ for AIA 171~\text{\AA}, and $\sim 120\,\mathrm{DN/s}$ for AIA 193~\text{\AA} , reflecting significant chromospheric, transition region, and coronal heating. 
However, when the \text{He~\textsc{i}}~10830~\text{\AA} intensity falls below this threshold (blue dashed line in Figure~\ref{fig:about_Im}~(G)-(H)), the scatter around the fitted line increases and the linear correlation with $\text{EUV}$ brightness weakens. This effect is particularly pronounced for $\text{EUV}$ bands corresponding to higher characteristic temperatures and atmospheric layers, as indicated by the scatter plots.



\begin{figure*}[htb!]
    \centering
    \includegraphics[width=0.66\linewidth]{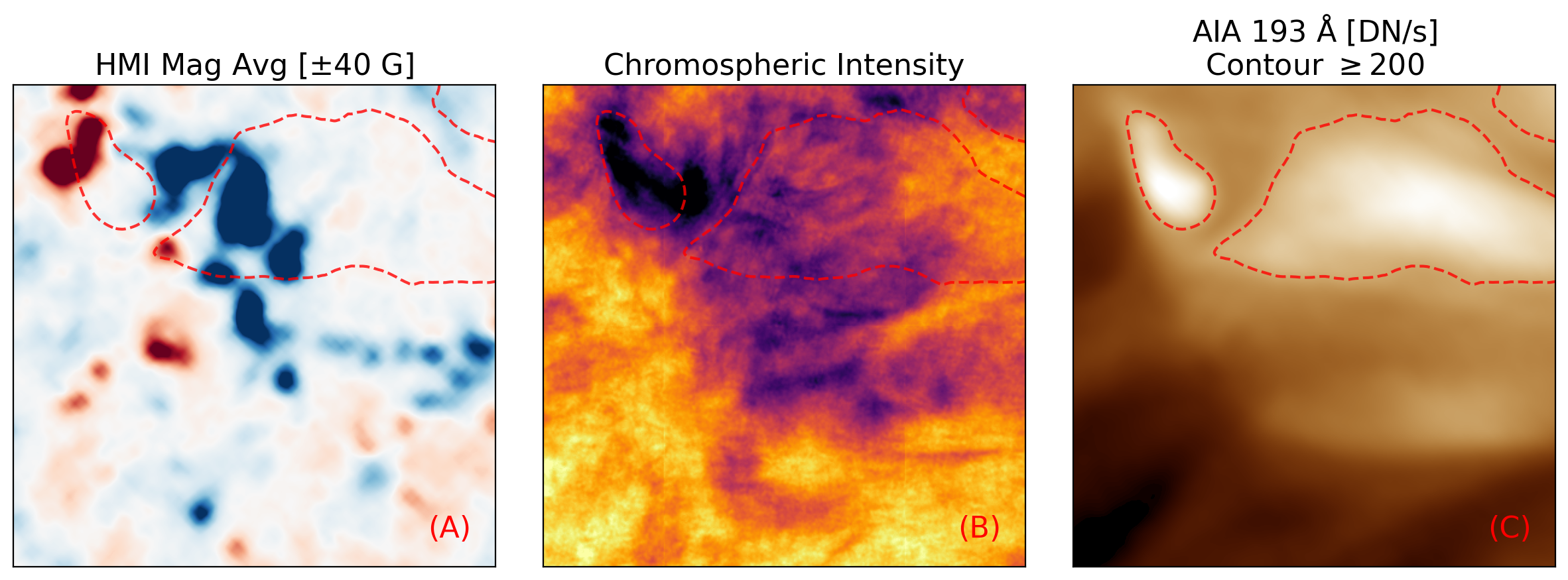}
    
    \caption{
        (A) The $\text{HMI}$ magnetogram;
        (B) The\text{He~\textsc{i}}~10830~\text{\AA} chromospheric signal;
        (C) The $\text{AIA 193\,\text{\AA}}$ image.
        The red dashed line outlines the regions where the $\text{193\,\text{\AA}}$ intensity is greater than 200 DN/s, representing the strongly heated regions observed in the $\text{193\,\text{\AA}}$ band.}
    \label{fig:higher atmosphere}
\end{figure*}

\subsection{Decoupled \text{He~\textsc{i}}~10830~\text{\AA}  Absorption and $\text{EUV}$ Brightening}
\label{subsec:Results3}

We first examine vertical differences in heating. The red dashed line in Figure~\ref{fig:higher atmosphere}~(C) outlines two bundles of coronal loops. The heated regions generally overlap with chromospheric absorption areas and strong magnetic field regions, but spatial inconsistencies also exist. 
The loop on the left connects both positive and negative strong magnetic regions and aligns well with the strong \text{He~\textsc{i}}~10830~\text{\AA} absorption area, whereas the brightening of the right loop clearly extends beyond the chromospheric absorption area and the strong magnetic regions.
This indicates layer-dependent heating: heating in the lower atmosphere is confined to strong magnetic regions, while coronal loops rooted on the solar surface transport energy along their length, heating the upper atmosphere. Consequently, the chromosphere beneath these surface-rooted loops is not necessarily a primary heating site, which explains why the $\text{AIA 193\,\text{\AA}}$ brightening can be strong while \text{He~\textsc{i}}~10830~\text{\AA}  absorption remains weak.

\begin{figure}[htb!]
    \centering
    \includegraphics[width=0.44\textwidth]{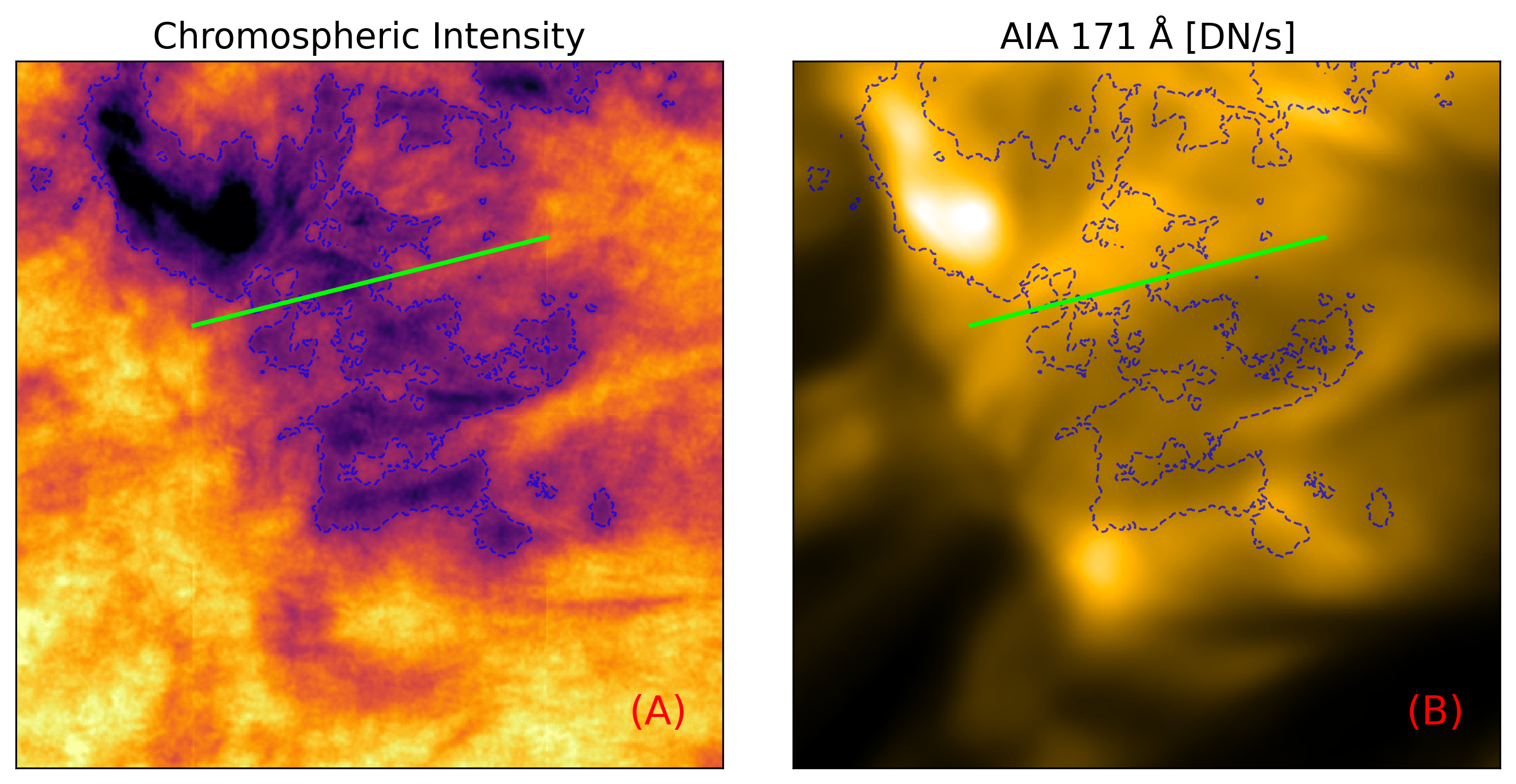}
    \caption{
        (A) The \text{He~\textsc{i}}~10830~\text{\AA}  chromospheric signal;
        (B) The $\text{171\,\text{\AA}}$ image.
        The area below the yellow-green line indicates our primary analysis region.}
    \label{fig:171_10830}
\end{figure}

We next examine the spatial structure of heating, as revealed by the \text{He~\textsc{i}}~10830~\text{\AA}  absorption and $\text{AIA 171\,\text{\AA}}$ emission. Although strong magnetic field regions generally correspond to stronger \text{He~\textsc{i}}~10830~\text{\AA} absorption, they do not necessarily exhibit strong $\text{AIA 171\,\text{\AA}}$ emission. This indicates that the significant enhancement of \text{He~\textsc{i}}~10830~\text{\AA} absorption is closely tied to the strong magnetic field, whereas the $\text{AIA 171\,\text{\AA}}$ emission may extend toward the edges of the strong magnetic field or along the neutral line.
As illustrated in the region to the lower right of the line segment in Figure~\ref{fig:171_10830}, the brightening region observed in the $\text{AIA 171\,\text{\AA}}$ band spatially overlaps with the strong absorption region of \text{He~\textsc{i}}~10830~\text{\AA} .
However, the strong $\text{AIA 171\,\text{\AA}}$ brightening region extends clearly beyond both the strong \text{He~\textsc{i}}~10830~\text{\AA} absorption region and the underlying strong magnetic field area, with its strong emission concentrated along the edges of the strong absorption region
That is, the strong $\text{AIA 171\,\text{\AA}}$ brightening region envelops the strong \text{He~\textsc{i}}~10830~\text{\AA}  absorption feature and its underlying strong magnetic area, yet does not cover them, indicating a clear separation of thermal structures.

\section{Discussion}\label{sec:Discussion}


This study develops a novel deep learning framework based on the $\text{TiO}$ cross-band learning technique. This framework achieves a precise and effective separation of the photospheric and chromospheric signals embedded within the \text{He~\textsc{i}}~10830~\text{\AA}  line core.
Consequently, this advanced method significantly enhanced the observation of chromospheric fine structures in the \text{He~\textsc{i}}~10830~\text{\AA}  band, yielding enhanced clarity and resolution.
Utilizing the results from this separation, we conducted a detailed analysis (Section $\text{\ref{sec:Results}}$) of the fine chromospheric activities associated with weak, small-scale coronal heating in the quiet Sun. 
Furthermore, we investigated the vertical energy coupling between the photospheric magnetic field and the overlying chromospheric and coronal activities, providing new observational constraints for understanding energy transfer in the solar atmosphere.



As detailed in Section~\ref{subsec:Results1}, we confirmed the ubiquitous vertical coupling in the quiet solar atmosphere. 
The \text{He~\textsc{i}}~10830~\text{\AA} and $\text{EUV}$ band intensities exhibit a widespread strong negative spatial correlation, with correlation coefficients $\text{R}$ ranging from $-0.84$ to $-0.66$. 
In comparison, the spatial correlations between the unseparated \text{He~\textsc{i}}~10830~\text{\AA} intensity and the EUV bands are notably weaker, yielding coefficients of $-0.72$ (AIA $304~\text{\AA}$), $-0.68$ (AIA $171~\text{\AA}$), and $-0.56$ (AIA $193~\text{\AA}$), respectively.
The reduction in correlation coefficients with elevated EUV formation heights and characteristic temperatures---as observed from the chromosphere and transition region (304\,\text{\AA}, He~\textsc{ii}, $\log T \approx 4.7$) up to the upper transition region and quiet corona (171\,\text{\AA}, Fe~\textsc{ix}, $\log T \approx 5.8$) and corona or hot flare plasma (193\,\text{\AA}, Fe~\textsc{xii}, $\log T \approx 6.1$; Fe~\textsc{xxiv}, $\log T \approx 7.3$) \citep{Cite16} , indicates that \text{He~\textsc{i}}~10830~\text{\AA} absorption correlates more strongly with heating in the lower atmospheric layers within quiet solar regions.
To interpret this correlation physically, we analyze the mechanism from the perspective of spectral line excitation. Specifically, heating processes in the chromosphere and corona excite orthohelium to the high-energy metastable state $\text{2s}\,^3\text{S}$. This excitation enhances the characteristic absorption of \text{He~\textsc{i}}~10830~\text{\AA}  photons via the $\text{2s}\,^3\text{S} \to \text{2p}\,^3\text{P}$ transition. Conversely, the same heating processes induce brightening in the EUV band\citep{Cite0_1, Cite0_2}.
This strong negative correlation strongly supports a tight vertical coupling between the chromosphere and corona, suggesting that the heating processes in these two atmospheric layers are closely linked. Although previous studies did not perform such precise quantitative fitting, our conclusion is consistent with the qualitative findings that $\text{EUV}$ band radiation is stronger in $\text{He\,\text{I}}~10830\,\text{\AA}$ absorption regions \citep{Cite11} and that these areas correspond to regions of coronal heating \citep{Cite4}.
\textbf{The superior spatial correlation between AIA 304~\text{\AA} and \text{He~\textsc{i}} 10830~\text{\AA} (compared to the AIA 171 ~\text{\AA} and 193 ~\text{\AA} bands) likely results from the combined contribution of the PR mechanism and collisional excitation.
According to the FAL-C model \citep{R2_1}, both \text{He~\textsc{i}} 10830~\text{\AA} and \text{He~\textsc{ii}} 304~\text{\AA} involve PR mechanism and collisional excitation, with the latter being the primary driver for \text{He~\textsc{ii}} 304~\text{\AA} in quiet Sun regions \citep{R2_2}. 
In quiet Sun regions, \text{He~\textsc{i}} 10830~\text{\AA} typically forms below 2,000 km , although it can extend up to 2,200 km with decreasing intensity. \citep{Cite0_5}, while \text{He~\textsc{ii}} 304~\text{\AA} can also form below this altitude \citep{R2_3}.
At 2,200~km, where the VAL-C model temperature reaches $\sim$24,000~K \citep{R2_4}, collisional excitation processes exist for both \text{He~\textsc{i}} 10830~\text{\AA} and \text{He~\textsc{ii}} 304~\text{\AA}.
This suggests that the observed spatial correlation results from the combined contribution of both PR mechanism and collisional processes. This effectively explains the stronger correlation between \text{He~\textsc{i}} 10830~\text{\AA} and AIA 304~\text{\AA} as compared to the AIA 193~\text{\AA} and 171~\text{\AA} bands. }


Analysis of the strong absorption regions (as discussed in Subsection $\text{\ref{subsec:Results2}}$) reveals that the spatial distribution of both chromospheric absorption and coronal brightening coincides with the strong magnetic field regions. This finding is consistent with the view that coronal heating is magnetically driven \citep{Cite5}.
It is noteworthy that strong chromospheric absorption does not necessarily correlate with the strongest coronal brightening. However, the $\text{EUV}$ intensity consistently exceeds a brightness threshold across all these strong absorption areas. This implies that the corona in these regions has been heated above a certain temperature.


Furthermore, observations of atmospheric layer inconsistency (as discussed in Subsection~\ref{subsec:Results3}) highlight the complexity of vertical energy transport. The coronal loop topology, specifically the $\text{AIA 193\,\text{\AA}}$ brightening regions extending above the non-absorbing chromospheric area, is direct evidence of energy layer decoupling. This confirms that the energy in this region is efficiently transported along magnetic field lines and dissipates primarily in the corona, rather than being fully consumed in the underlying chromosphere.


Our observations reveal that the most intense $\text{AIA 171\,\text{\AA}}$ emission regions are distributed at the edges of the strong \text{He~\textsc{i}}~10830~\text{\AA} absorption areas and unipolar magnetic fields, but they do not completely overlap with the strong absorption region. This spatial difference suggests that the heating mechanisms at different atmospheric heights may possess fundamental disparities \citep{DISCUSSION_1}. 
Firstly, a significant co-spatial relationship exists between the strong unipolar magnetic field regions and the intense \text{He~\textsc{i}}~10830~\text{\AA} absorption signals.
It is inferred that the chromospheric heating process in these areas likely originates from the random magnetic field motion mechanisms proposed by Parker \citep{DISCUSSION_2}.
However, the coronal brightness exhibits higher intensity within the mixed-polarity edge regions. In these areas, the photospheric magnetic flux cancellation and reconnection likely constitute the dominant mechanism for transporting energy and mass into the corona \citep{DISCUSSION_3}. 
This process effectively accounts for the enhanced $\text{EUV}$ brightness observed at the edges compared to the unipolar interiors. Furthermore, the continuous emergence and cancellation of magnetic fields in these edge areas promote reconnection with overlying atmospheric fields \citep{DISCUSSION_4}, resulting in sustained energy release into the upper atmosphere. 
This mechanism corroborates the stronger edge $\text{EUV}$ brightening relative to the unipolar region interiors. In summary, the observed spatial non-overlap between the strong chromospheric absorption areas and the coronal brightening areas is collectively explained by these distinct magnetically driven heating mechanisms.


Despite the considerable success of our deep learning approach in separating the \text{He~\textsc{i}}~10830~\text{\AA}  signal, several methodological and data acquisition limitations remain. 
These challenges primarily arise from the inherently low signal-to-noise ratio of the \text{He~\textsc{i}}~10830~\text{\AA}  band and the fact that the independent reconstruction of $\text{TiO}$ and \text{He~\textsc{i}}~10830~\text{\AA}  may disrupt the cross-band intensity correlation. Moreover, during the early phase of this study, the $\text{NVST}$ infrared channel had only recently begun operation, making it difficult to obtain high-quality datasets for model training. As a result, the model's training and test sets contain only a very low proportion of photospheric samples from sunspot regions, with a complete exclusion of those from the solar limb.
These limitations significantly compromise the generalization capability of the AR model ($\text{Model A}$). Predictions of key parameters such as sunspots, intergranular lane depth and mean intensity in sunspot regions are biased.
Furthermore, the exponential absorption model underlying the AR model is not universally applicable to all active region scenarios.
Consequently, the application of $\text{Model A}$ is highly condition-dependent and subject to limitations. In contrast, the QS model ($\text{Model B}$) was specifically designed for the relatively uniform structural characteristics of the QS, resulting in fewer generalization issues. Therefore, accurately assessing and defining the practical scope of application for both models is essential.

\section{Conclusion}
\label{sec:Conclusion}

This study presents a novel deep learning framework based on $\text{TiO}$ cross-band learning. Utilizing this technique, we achieved a precise and effective separation of the photospheric and chromospheric contributions within the \text{He~\textsc{i}}~10830~\text{\AA} signal. 
This separation facilitated the generation of high-contrast images of chromospheric fine structures, allowing the clear observation of features previously obscured by strong photospheric background emission.
Applying this method, we successfully observed chromospheric fine structures associated with weak, small-scale coronal heating processes prevalent in the $\text{QS}$. Our subsequent analysis provided direct observational evidence of energy coupling between the chromosphere and the overlying atmosphere.
Our research reveals several specific findings regarding vertical energy transport in the solar atmosphere.

Firstly, we observed a significant vertical coupling between the \text{He~\textsc{i}} 10830~\text{\AA} and \text{EUV} intensities(AIA 304~\text{\AA}, 171~\text{\AA}, and 193~\text{\AA}), which are negatively correlated. This coupling is most pronounced in the AIA 304~\text{\AA} band.
Additionally, we found that within the strong absorption regions, the $\text{EUV}$ band brightness does not necessarily reach its peak intensity, exhibiting a degree of non-linear relationship; however, it is notable that the $\text{EUV}$ intensity within these strong absorption areas is consistently heated to a certain brightness threshold.
Furthermore, observations of the $\text{AIA 193\,\AA}$ brightening regions extending above the non-absorbing chromospheric area provide strong evidence that energy is efficiently transported along magnetic field lines  and dissipates primarily in the upper atmosphere, resulting in layer decoupling. 
We observed that the \text{He~\textsc{i}}~10830~\text{\AA}  strong absorption feature is intense in the unipolar regions of the $\text{QS}$, but the $\text{AIA 171\,\AA}$ brightening is stronger in the mixed-polarity regions. 
This spatial separation confirms that the magnetic field topology strongly constrains the location of heating footpoints. It further demonstrates that the coronal heating mechanisms differ between mixed-polarity regions and unipolar regions in the quiet Sun.



\begin{acknowledgments}
The authors thank the team of the New Vacuum Solar Telescope (NVST), and NASA's Solar Dynamics Observatory (SDO) for the data used in this paper.
We also appreciate all the help from the colleagues in the laboratory team.
This work was primarily supported by the National Key R\&D Program of China under grant No. 2024YFA1612003, and performed by the Yunnan Observatories, Chinese Academy of Sciences (CAS).
Additionally, this work was supported by the National Science Foundation of China under grants 12273106 and 12273108, the $\text{CAS}$ “Light of West China” Program, and the “Yunnan Revitalization Talent Support Program” Innovation Team Project 202405AS350012.
\end{acknowledgments}

\facilities{NVST, SDO(AIA), SDO(HMI)}
\software{astropy \citep{2013A&A...558A..33A,2018AJ....156..123A,2022ApJ...935..167A}, 
          PyTorch \citep{2019arXiv191201703P}}

\appendix
\section{Network Architectures}
\label{appendix:models}

This appendix provides a detailed comparison of the two neural network architectures.

\begin{figure}[htbp]
    \centering
    \includegraphics[width=0.8\textwidth]{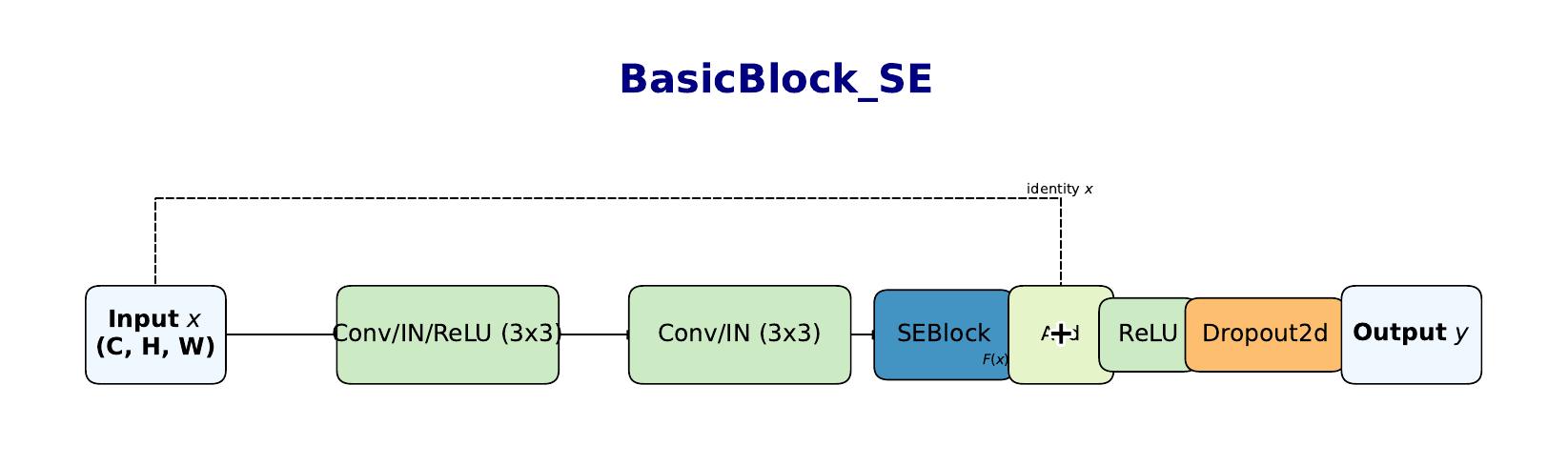} 
    \caption{Basic residual block incorporating Squeeze-and-Excitation (SE) attention mechanism, used in Model A.}
    \label{fig:basic_block}
\end{figure}

\begin{figure}[htbp]
    \centering
    \includegraphics[width=0.98\textwidth]{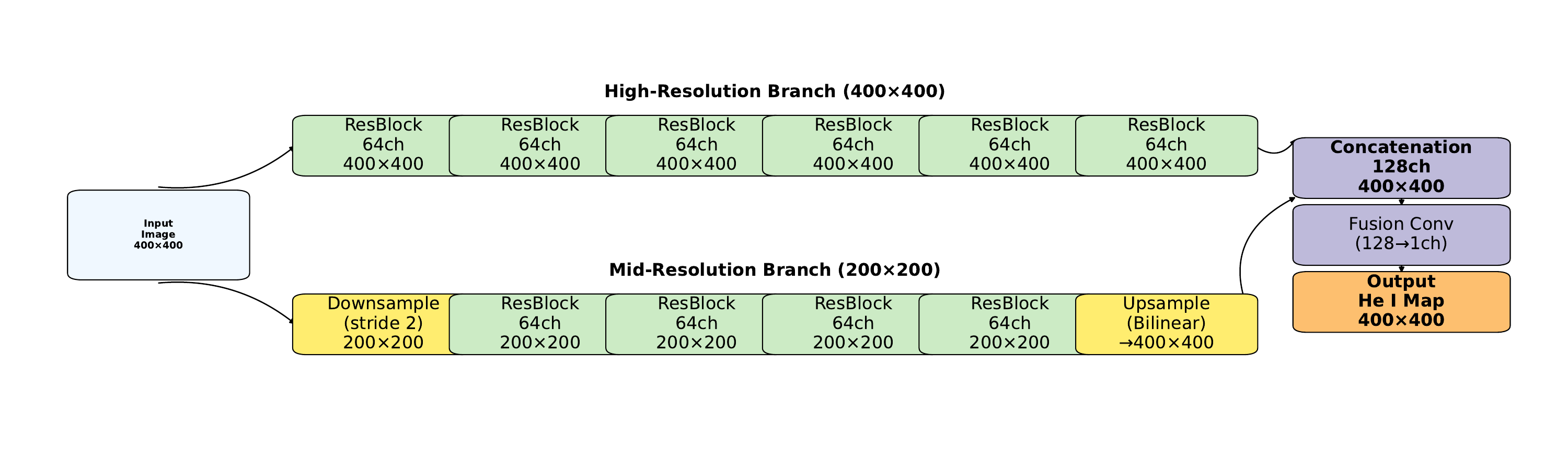}
    \caption{Detailed architecture of the Model A. The model maps TiO input to 10829 \AA\ to leverage the photospheric information of 10829 \AA\ as a proxy for the 10830 \AA\ photosphere.}
    \label{fig:model_a_full}
\end{figure}

\begin{figure}[htbp]
    \centering
    \includegraphics[width=0.8\textwidth]{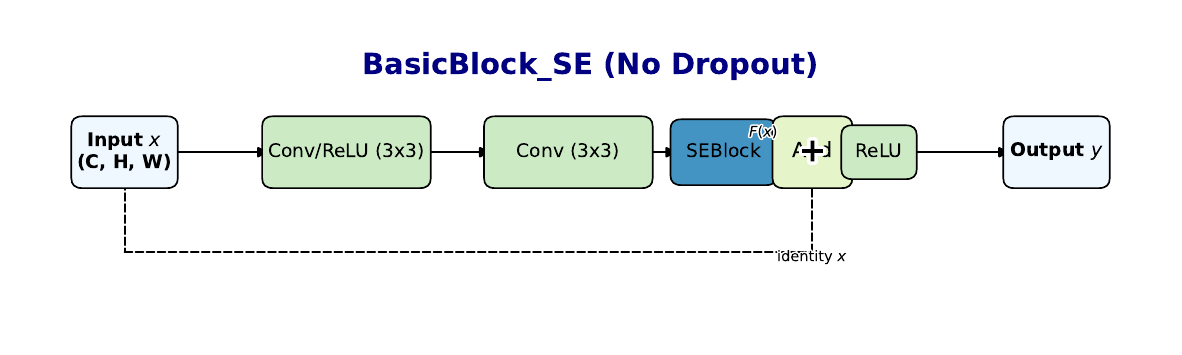} 
    \caption{Modified residual block for Model B, with BN and Dropout removed to better fit continuous observations of a specific region rather than cross-date generalization.}
    \label{fig:basic_block_b}
\end{figure}

\begin{figure}[htbp]
    \centering
    \includegraphics[width=0.98\textwidth]{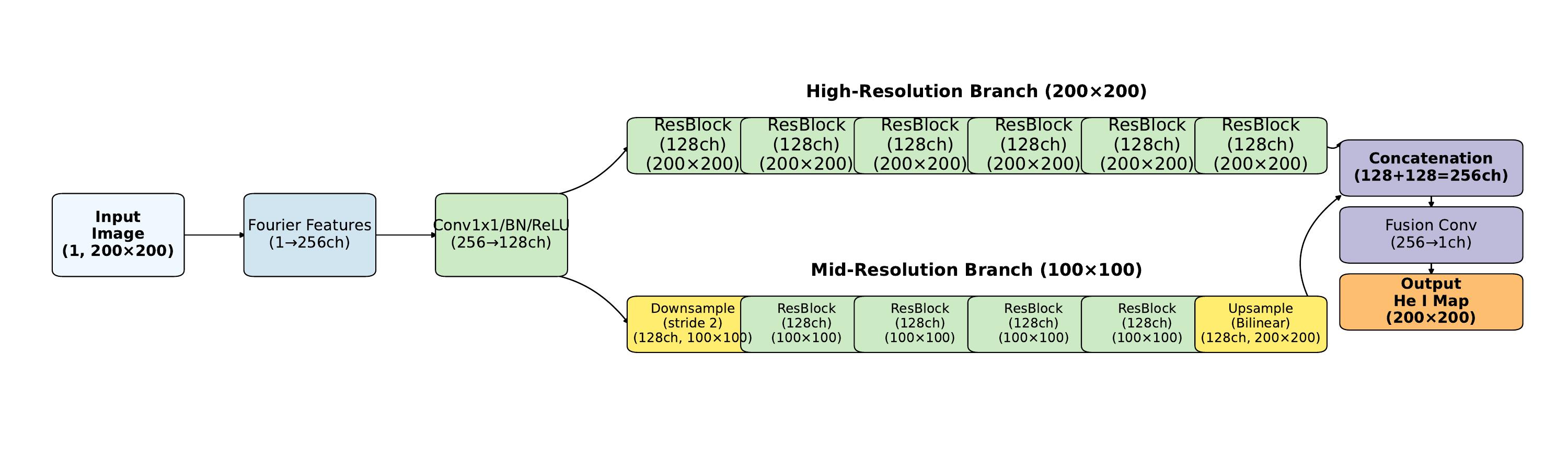}
    \caption{Detailed architecture of the Model B for Quiet Sun (QS) regions. The model directly maps TiO to 10830 \text{\AA} by treating the weak and rapidly varying chromospheric signals as residuals, thereby establishing a mapping relationship between TiO and the photospheric information of 10830 \text{\AA}.}
    \label{fig:model_b_full}
\end{figure}

\clearpage

\bibliography{sample701}{}
\bibliographystyle{aasjournalv7}
\end{document}